\documentclass[aps,amsmath,amssymb,prl,preprint,groupedaddress,nofootinbib,a4paper]{revtex4-1}
\pdfoutput=1
\usepackage[DIV13]{typearea}
\usepackage{amsmath,cancel,bbold}
\usepackage{amsfonts}
\usepackage{mathrsfs}
\usepackage{amssymb}
\usepackage[usenames,dvipsnames]{color}
\usepackage{feynmp}
\usepackage{multirow}
\usepackage{array}
\usepackage{slashed}
\usepackage{units}
\usepackage{bbm}
\usepackage[left=.9in, right=.9in]{geometry}
\usepackage[titletoc]{appendix}
\usepackage{hyperref}
\usepackage{colortbl,framed,feynmp,graphicx}
\usepackage[usenames,dvipsnames]{xcolor}
\usepackage[nottoc,numbib]{tocbibind}

\newcommand\sss{\scriptscriptstyle}

\newcommand{\cW}{c_{\sss W}}

\def\bsp#1\esp{\begin{split}#1\end{split}}
\def\bpm{\begin{pmatrix}}
\def\epm{\end{pmatrix}}

\newcommand{\g}{\gamma}

\renewcommand{\to}{\rightarrow}

\newcommand{\de}{\partial}

\newcommand{\beq}{\begin{equation}}
\newcommand{\eeq}{\end{equation}}
\renewcommand{\[}{\begin{equation}}
\renewcommand{\]}{\end{equation}}
\newcommand{\bmat}{\begin{pmatrix}}
\newcommand{\emat}{\end{pmatrix}}

\newcommand{\bea}{\begin{eqnarray}}  
\newcommand{\eea}{\end{eqnarray}}  

\newcommand{\Mp}{M_{\rm p}}

\newcommand{\Mev}{\,{\rm MeV}}
\newcommand{\phimin}{v_\phi}
\newcommand{\mphi}{m_\phi}
\newcommand{\mchi}{m_\chi}
\newcommand{\qtc}{g^2}
\newcommand{\tri}{\sigma}
\newcommand{\cn}{\text{cos}}
\newcommand{\sn}{\text{sin}}
\newcommand{\tn}{\text{tan}}
\newcommand{\p}{\prime}

\begin{document}
\title{Reheating with a Composite Higgs}
\author{Djuna Croon, Ver\'onica Sanz and Ewan R. M. Tarrant}
\affiliation{University of Sussex}

\date{\today}

\begin{abstract}
The flatness of the inflaton potential and lightness of the Higgs could have the common origin of the breaking of a global symmetry. This scenario provides a unified framework of Goldstone Inflation and Composite Higgs, where the inflaton and the Higgs both have a pseudo--Goldstone boson nature. The inflaton reheats the Universe via decays to the Higgs and subsequent secondary production of other SM particles via the top and massive vector bosons.  We find that inflationary predictions and perturbative reheating conditions are consistent with CMB data for sub--Planckian values of the fields, as well as opening up the possibility of inflation at the TeV scale. We explore this exciting possibility, leading to an interplay between collider data cosmological constraints.
\end{abstract}

\maketitle
\tableofcontents

\pagebreak

%%%%%%%%%%%%%%%%%%%%%%%%%%%%%%%%%%%%%%%%%%%%
\section{Introduction}

Scalar fields are popular protagonists in cosmological theories. They play chief roles in the leading paradigms for important events, such as inflation and electroweak symmetry breaking. However, it has been long known that fundamental scalars suffer radiative hierarchy problems: for theory to match observations, one requires an unnatural cancelation of UV corrections. In inflation, this radiative instability can be quantified by the tension between the Lyth bound~\cite{Lyth:1996im} on the slow roll phase of the field, pushing towards $\Delta \phi > M_p$, and the measurement of CMB anisotropies, which indicate $\Lambda_{inf} \lesssim 10^{15}\, \text{GeV}$. For electroweak symmetry breaking (EWSB), one usually considers the large separation of scales between the Higgs mass and the Planck scale as an illustration, as the latter is where the theory should be cut off for an elementary Higgs. 

Here we will discuss the appeal of pseudo--Goldstone bosons (pGBs) for the dynamical generation of scales in both paradigms. The realisation that Goldstone bosons can solve hierarchy problems is not new: for EWSB, there is popular branch of model building that goes by Composite Higgs theory which postulates a new strongly coupled sector of which the Higgs is a bound state \cite{Agashe:2004rs} (for a review see \cite{Bellazzini:2014yua}). The effective theory then has a cut-off, such that the Higgs mass is not sensitive to effects above the compositeness scale. 

Likewise, in inflationary model building ``Natural Inflation'' provides an inflaton candidate protected from UV corrections using essentially the same mechanism with an axionic GB \cite{Freese:1990rb}. Alas, vanilla Natural Inflation requires trans--Planckian scales to predict the measured Cosmic Microwave Background (CMB) spectrum and thus has questionable value as a valid effective theory.\footnote{There have been several proposals to explain the trans--Planckian decay constant while maintaining the simple potential and the explanatory power of the model. Among these are Extra--Natural inflation~\cite{ArkaniHamed:2003wu}, hybrid axion models~\cite{Linde:1993cn, Kim:2004rp}, N-flation~\cite{Dimopoulos:2005ac, Copeland:1999cs}, axion monodromy~\cite{Silverstein:2008sg} and other pseudo-natural inflation models in Supersymmetry~\cite{ArkaniHamed:2003mz}. } 
In \cite{Croon:2014dma} the idea of a pGB inflaton was generalised, and it was shown there and in \cite{Croon:2015fza} that different models may realise inflation compatible with data from the Cosmic Microwave Background (CMB) without the issues that the original Natural Inflation has. 

In this paper we will show how both mechanisms can be unified, thus realizing radiative stability for both models in a single simple set--up. We will explore the minimal symmetry breaking pattern that realises a Higgs $SU(2)$ doublet and an inflaton singlet. We discuss both the generation of an inflaton potential and reheating in this model. Interestingly, both can be fully perturbative processes. The inflationary predictions are shown to be compatible with the latest CMB data by Planck \cite{Ade:2015lrj} without the necessity of introducing trans-Planckian scales in the effective theory. After inflation the inflaton decays into Higgs bosons, which subsequently decay into the Standard Model particles. 
Importantly, we find that the question if reheating can take place perturbatively crucially depends on the CP assignment in the model. 

\begin{figure}[ht] \label{scales}
\centering
  \includegraphics[width= 300pt]{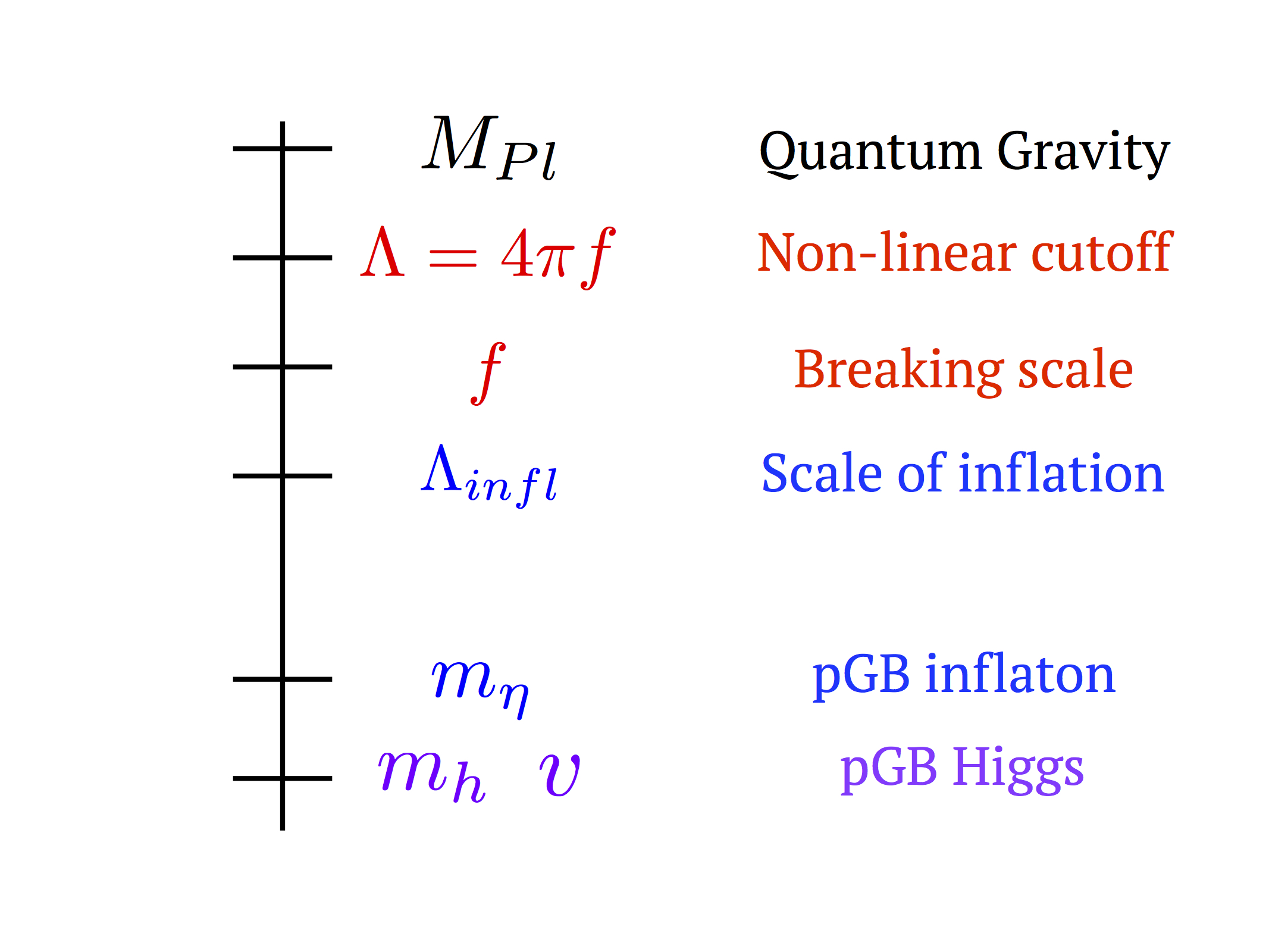}
  \caption{{\it Relevant scales:} pseudo-Goldstone bosons naturally realise mass hierarchies. CMB data and constraints on perturbative reheating allow us to relate the complete spectrum to the symmetry breaking scale $f$ and the Planck scale $M_p$. }
\end{figure}
We will finish by showing how the model naturally connects to electroweak physics. The inflaton mass and couplings to the Higgs could be of the same order, leading to the possibility of looking for the inflaton through their mixing with the Higgs. 

In Fig.~\ref{scales} we show a graphic of the relevant scales in our model. The global symmetry is broken at the scale $f$, which is below the Planck scale at which we expect a UV completion in the form of a theory of quantum gravity. The scale of inflation is then expected to be parametrically smaller than $f$, as we will show. The Coleman Weinberg masses of the goldstone boson inflaton and Higgs are fixed by CMB and electroweak data respectively. 
Likewise, the values of the coefficients of the (self-) couplings in the potential can be fixed in light of the data, modulo the scale of inflation. This is a free parameter in our model. As usual for slow roll inflation, it is most naturally found around the GUT scale ($10^{15} \, \text{GeV}$), but can be as low as $\sim 10^5\, \text{GeV}$ if one allows for a degree of tuning.

Finally we would like to highlight some recent developments that may be of interest to the reader. In \cite{Graham:2015cka} a dynamical solution to the electroweak hierarchy problem was proposed, in terms of a Higgs boson coupling to an inflaton and an axion--like field. Although critics have pointed out several shortcomings, among which the necessity of a very large number of e--foldings and the low cut--off (which makes one arguably expect new physics around the EW scale) \cite{Espinosa:2015eda}, the scanning mechanism is a new facet worth investigating. As the model behind the mechanism bares similarities with our set--up, it seems like a worthwhile exercise to look for a realisation in the present context.    
A second recent result that is interesting in the present context is the observation in \cite{Gross:2015bea} that the Higgs-inflaton coupling $c_4 h^2 \eta^2$ may drastically alter the Higgs dynamics in the Early Universe, thereby stabilising the electroweak vacuum. As we will see the coupling $c_4$ will automatically be present in our model.

\iffalse
Then it was also recently found that the presence of a singlet may affect decay patterns of top partners. [http://arxiv.org/abs/1506.05110] The analysis rests on a light inflaton, such that the decay $T \rightarrow \eta t$ is opened up. The authors argue that this will weaken the experimental bounds on top partner masses.  
\fi

%%%%%%%%%%%%%%%%%%%%%%%%%%%%%%%%%%%%%%%%%%%%
\section{The Lagrangian of the Higgs and the Inflaton}
\subsection{Inflaton--Higgs couplings for perturbative reheating}
The condition that the inflaton field must decay completely into relativistic particles to complete the reheating process dictates the interaction structure in a successful theory of inflation. After the end of inflation, the inflaton field $\eta$ begins to oscillate about the minimum of its potential with amplitude $\Phi(t)$. The universe is completely dominated by the zero--mode, $\langle\eta(t)\rangle$, which may be interpreted as a condensate of non--relativistic zero--momentum $\eta$--particles of mass $m_\eta$. The condensate oscillation amplitude decays as $\Phi(t) \sim t^{-1}$ due to the Hubble expansion and due to interactions with the higgs field. Trilinear couplings, $\frac12\tri\eta h^2$, and quartic couplings, $\frac12\qtc\eta^2 h^2$, with the higgs are to be expected on fairly general grounds, as we argue in the following section. As we will show in section~\ref{sec:reheating}, provided that the coupling constants $\tri,\qtc$ and the amplitude $\Phi(t)$ are small enough such that non--perturbative particle production processes are absent, the energy loss experienced by the condensate can be described by the Boltzmann equation
\beq
\frac{\rm d}{{\rm d}t} \left(a^3\rho_\eta \right) = 
- \frac{\tri^2\Phi_0^2m_\eta}{64\pi} 
-  \frac{g^4\Phi_0^4m_\eta}{128\pi a^3}\,,
\eeq
where $a$ is the scale factor and $\Phi_0$ is the initial amplitude of the inflaton oscillations at the start of reheating. The contribution from the quartic interaction decreases as $a^{-3}\sim t^{-2}$, which, as is well known~\cite{Kofman:1997yn,Podolsky:2005bw,Braden:2010wd}, poses a major problem for theories which do not contain a trilinear interaction.  Specifically, since the Hubble rate decreases as $H\sim a^{-3/2}\sim t^{-1}$, volume dilution due to the Hubble expansion takes place faster than the annihilation process $\phi\phi\to\chi\chi$ can drain energy from the condensate and so reheating never completes. In order to successfully reheat the universe, a trilinear coupling must be present. We will use this result as a guiding principle when constructing the Lagrangian for the composite Higgs model.

\subsection{Symmetry breaking: the minimal coset}
The inflaton and Higgs corresponds to five scalar degrees of freedom which could come from the breaking of $SO(6)$ to  $SO(5)$ or, equivalently $SU(4)$ to $Sp(4)$. This breaking pattern is very popular in building models of Composite Higgs, as it preserves custodial symmetry.  

The breaking gives rise to five Goldstone bosons, transforming as a $\mathbf{5}$ of $SO(5)$.
The most general vacuum which breaks $SO(6)\rightarrow SO(5) \sim SU(4) \rightarrow Sp(4)$ as shown in Ref.~\cite{1001.1361} is given by\footnote{The discussion in Ref.~\cite{1001.1361} assumes the presence of CP conserving vacua, as well as CP breaking vacua, such that the Pfaffian of the inflaton is real.}
\bea \Sigma_0 = \left(
\begin{array}{cccc}
 0 & e^{i \alpha } \cos (\theta ) & \sin (\theta ) & 0 \\
 -e^{i \alpha } \cos (\theta ) & 0 & 0 & \sin (\theta ) \\
 -\sin (\theta ) & 0 & 0 & -e^{-i \alpha } \cos (\theta ) \\
 0 & -\sin (\theta ) & e^{-i \alpha } \cos (\theta ) & 0 \\
\end{array}
\right) \eea
where $\alpha$ and $\theta$ are real angles. One recovers a well known choice of vacuum in Composite Higgs models~\cite{0902.1483} in the limit $\alpha \rightarrow \,\,\text{mod}(\pi)$ and $\theta \rightarrow \,\,\text{mod}(\pi)$.

In fact, the vacuum in which we have $\theta = \text{mod}(\pi)$ the vacuum has an enhanced custodial symmetry~\cite{1001.1361}, as in this case the unbroken generators generate $ \text{SU(2)} \times \text{SU(2)} \subset  \text{Sp(4)} $. Likewise, the limit $\alpha = \text{mod}(\pi)$ parametrises the conservation of CP by the vacuum. 

One can then parametrise the Goldstone bosons via the field $\Sigma(x)$, 
\bea \Sigma(x) = e^{i \Pi^{a}(x) T_\bot^{a} /\sqrt{2} f} \Sigma_0 \ , \eea 
where $\Pi^{a}(x) $ are the Goldstone fields with decay constant $f$, corresponding to the broken $SO(6) \cong SU(4)$ generators $T_\bot^{a}$. A linear combination of three of the Goldstone fields is eaten by the Standard Model gauge fields such that the corresponding generators can be recognised as their longitudinal components. 
The two remaining Goldstone bosons remain in the spectrum as massless scalar fields and couple via the broken generators $T_\bot^4$ and $T_\bot^5$:\footnote{Here we use generalized expressions from Ref.~\cite{1001.1361}; obtained by assuming the general vacuum (Eq. A. 17) in the rotation Eq. B. 25.} 
\bea T_\bot^4 = \left(\begin{array}{cc}
 0   & \sigma_2  \\
 \sigma_2 &  0 \\
\end{array} \right), \,\,\,\,\,\,\,\,\,\,\,\,\,\,\,\,\,\,\,\,\,\,\,\,\, T_\bot^5 = \left(\begin{array}{cc}
 c_\theta e^{i \alpha} \mathbb{1}_2  & -i s_\theta \sigma_2  \\
 i s_\theta \sigma_2 &  c_\theta e^{i \alpha} \mathbb{1}_2 \\
\end{array} \right). \eea
Expanding the matrix exponential, we obtain \iffalse (see appendix in Ref.~\cite{1001.1361}), setting the unphysical ("eaten") fields to zero and expanding we have: 
$$ \Sigma(x) = \Pi^{\hat{a}}  T^{\hat{a}} \Sigma_0 =   \left(
\begin{array}{cccc}
 \cos (\theta ) \eta   & 0 & 0 & -i h -\sin (\theta ) \eta   \\
 0 & \cos (\theta ) \eta   & i h +\sin (\theta ) \eta   & 0 \\
 0 & \sin (\theta ) \eta  -i h  & -\cos (\theta ) \eta   & 0 \\
 i h -\sin (\theta ) \eta   & 0 & 0 & -\cos (\theta ) \eta   \\
\end{array}
\right) \Sigma_0$$
Such that \fi
\bea  \Sigma(x) =    \left(
\begin{array}{cccc}
 c_{\pi }+\frac{\sqrt{2} f i e^{i \alpha} c_{\theta } s_{\pi } \eta  }{\sqrt{\pi _a^2}} & 0 & 0 & \frac{i \sqrt{2} f s_{\pi } \left(-i h-s_{\theta } \eta  \right)}{\sqrt{\pi _a^2}} \\
 0 & c_{\pi }+\frac{\sqrt{2} f i e^{i \alpha} c_{\theta } s_{\pi } \eta  }{\sqrt{\pi _a^2}} & \frac{i \sqrt{2} f s_{\pi } \left(i h+s_{\theta } \eta  \right)}{\sqrt{\pi _a^2}} & 0 \\
 0 & \frac{i \sqrt{2} f s_{\pi } \left(s_{\theta } \eta  -i h \right)}{\sqrt{\pi _a^2}} & c_{\pi }-\frac{i \sqrt{2} f e^{i \alpha} c_{\theta } s_{\pi } \eta  }{\sqrt{\pi _a^2}} & 0 \\
 \frac{i \sqrt{2} f s_{\pi } \left(i h -s_{\theta } \eta  \right)}{\sqrt{\pi _a^2}} & 0 & 0 & c_{\pi }-\frac{i \sqrt{2} f e^{i \alpha} c_{\theta } s_{\pi } \eta  }{\sqrt{\pi _a^2}} \\
\end{array}
\right) \Sigma_0 \eea
where we have suppressed space-time dependence of the fields $h = h(x)$ and $\eta = \eta(x)$, and where we use the shorthands,
\begin{alignat}{3}\notag h(x)^2+\eta (x)^2 = \pi _a^2\,\,\,\,\,\,\,\text{ and }  \,\,\,\,\,\,\, s_{\pi } &= \sin \left(\frac{\sqrt{\pi _a^2}}{\sqrt{2} f}\right) \,\,\text{,}  \,\,\,\,\,\,\, && c_{\pi } = \cos \left(\frac{\sqrt{\pi _a^2}}{\sqrt{2} f}\right) \\
s_\theta &= \sin(\theta) \,\,\text{,}  \,\,\,\,\,\,\, && c_\theta = \cos(\theta) \,\,\,\,\,\,\,.\end{alignat}

We will further assume that gauging the theory breaks $SU(4)$ to the Standard Model group\footnote{Here we do not address the colour group $SU(3)_c$.} $SU(2)_L \times U(1)_Y$ and $U(1)_{\eta}$. This latter shift symmetry for $\eta$ will assure that it does not get a potential from gauge bosons.  
Then the kinetic term becomes,
\begin{align} \notag \frac{f^2}{8}\text{Tr} |D_\mu \Sigma |^2  = \frac{1}{2} \frac{ \left(\eta \partial_\mu h - h  \partial_\mu \eta \right)^2}{h ^2+\eta ^2} + \frac{g^2 }{4} h^2   \left( W^+_{\mu}W^{- \mu} + \frac{1}{\cos^2\theta_w}Z_\mu Z^\mu\right)  \\ 
\approx  \frac{1}{2}  (\partial_\mu h)^2+\frac{1}{2}  (\partial_\mu \eta)^2 + \frac{1}{2} \frac{ \left(h  \partial_\mu h +\eta \partial_\mu \eta \right)^2}{1- h ^2-\eta ^2} + \frac{g^2 }{4} h^2  \left( W^+_{\mu}W^{- \mu} + \frac{1}{\cos^2\theta_w}Z_\mu Z^\mu\right) \label{eq:Lkin} \end{align}
where the following field redefinitions are made:
\bea
\begin{array}{ll}
h^2 s_\pi^2 f^2/ \pi _a^2 \rightarrow h^2 &  \eta^2 s_\pi^2 f^2/ \pi _a^2 \rightarrow \eta^2  \\ 
(\partial_\mu h \, s_\pi f/ \sqrt{\pi _a^2})^2 \rightarrow (\partial_\mu h)^2 &(\partial_\mu \eta \,  s_\pi f/ \sqrt{\pi _a^2})^2  \rightarrow (\partial_\mu\eta)^2 \end{array}\eea
corresponding to dropping the operators with more than four powers in the field (they will be effectively suppressed by $f$). For the sigma model, there is an equivalence between the original and rotated fields. However, the rotated fields couple to gauge bosons as in \eqref{eq:Lkin} and are as such the physically relevant choice. 

At this level, the $\eta$ and $h$ fields are true Goldstone bosons. (Small) explicit breaking of the symmetry will generate a Coleman-Weinberg contributions to the scalar potential, via gauge and Yukawa interactions.  This potential accounts, then, for resummations of loops of gauge bosons and fermions. Rather than considering the fully generic case, we can use the information from the previous section as prior information about what a Lagrangian which gives perturbative reheating will look like. In particular, the necessity of terms with odd powers of the singlet $\eta$ in the scalar potential implies that the singlet $\eta$ has specific transformation properties under CP that differ from the Composite Higgs model. This can be understood in the following way: if we for a moment assume that CP is unbroken, we can set $\alpha = 0$. As we will see, the way we parametrise the coupling between $\eta$ and (Dirac) fermions can schematically be written as 
\bea\label{CPferm} \eta \bar{F} ( c_{even} + i c_{odd} \gamma_5 ) F \eea
Clearly, for $c_{odd} = 0$, $\eta$ behaves as a scalar, such that the trilinear interaction $\eta h^2$ is allowed by the symmetry. However in the Composite Higgs case ($c_{odd} \neq 0$) where $\eta$ behaves as a (partial) pseudo-scalar, the term $\eta h^2$ breaks CP. 

In contrast, the breaking of the enhanced custodial symmetry by taking $\theta \neq 0$ does not have such a direct impact on the predictions for perturbative reheating. It is expected to give rise to mass mixing, i.e. terms of the form $V \ni c_i \,\eta \, h$. Deviations from custodial symmetry in the Higgs sector are rather constrained by low-energy data and it will therefore be practical to assume $\theta = 0$ in the following. This choice corresponds to identifying the Higgs with the bi-doublet under the subgroup $SO(4) \cong SU(2)_L \times SU(2)_R$, and $\eta$ with the singlet: $\mathbf{1} \oplus \mathbf{4} = \mathbf{(1,1)} \oplus \mathbf{(2,2)}$.

As the scalar $\eta$ does not couple to the $SU(2)_L$ gauge group, see Eq.~\ref{eq:Lkin}, couplings to gauge bosons do not help with generating a cubic term. The difference in dynamics between the different vacua has to come from the couplings to fermions. 

As an example, we implement the fermions in a \textbf{6} of SU(4) (corresponding to the vector representation of SO(6)).  Other options for fermion representations, such \textbf{4} and the \textbf{10}, have their own difficulties to address~\cite{0902.1483}. %We will address this choice briefly later on, because as we will see we will have to introduce a small CP breaking in the model to make it a viable candidate for perturbative reheating. 

The \textbf{6} of SU(4) decomposes as $(\bf{2},\bf{2}) \oplus (\bf{1},\bf{1}) \oplus (\bf{1},\bf{1})$ under $\text{SU(2)}_L \times \text{SU(2)}_R$, such that we can implement the fermions as~\cite{0902.1483}
\begin{subequations}
\bea \Psi_q = \frac{1}{2} \left( \begin{array}{cc}
0 & Q \\
-Q^T & 0 \end{array} \right) \,\,\,\,\,\,\,\,\,\,\,\,\,\,\,\,\,\, \Psi_u = \Psi_u^+ +   \Psi_u^-\,\,\,\,\,\,\,\,\,\,\,\,\,\,\,\,\,\, \Psi_u^{\pm} = \frac{1}{2} \left( \begin{array}{cc}
\pm U & 0 \\
0 & U \end{array} \right) \eea
\bea \Psi_{q'} = \frac{1}{2} \left( \begin{array}{cc}
0 & Q' \\
-Q'^T & 0 \end{array} \right) \,\,\,\,\,\,\,\,\,\,\,\,\,\,\,\,\,\, \Psi_d = \Psi_d^+ + \epsilon_d \Psi_d^-\,\,\,\,\,\,\,\,\,\,\,\,\,\,\,\,\,\, \Psi_d^{\pm} = \frac{1}{2} \left( \begin{array}{cc}
\pm D & 0 \\
0 & D \end{array} \right) \eea
\end{subequations}
where $Q = (0, q_L)$, $Q' = (q_L,0)$, $U= u_R i \sigma_2$ and $D= d_R i \sigma_2$. The $\epsilon_{u,d}$ are complex free parameters defining the embedding of the quarks into the singlets, and consecutively the CP-assignment of $\eta$. In the limit $|\epsilon_{u,d}| = 1$ the fermions have definite charges under $U(1)_\eta$ and it is therefore expected that $\eta$ is massless. 

The coupling of $\Sigma$ to fermions will be of the form
\begin{align}\notag \mathcal{L}_{eff} = &\sum_{r =q,u,q',d} \left[ \Pi^r_{0} \text{Tr}[\bar\Psi_r \slashed{p}  \Psi_r] + \,\Pi^r_{1} \text{Tr}\left[\bar\Psi_r \Sigma \right]  \slashed{p} \text{Tr}[\Psi_r \Sigma^{\dagger}] \right] \\
&+  M_{u} \text{Tr}\left[ \bar\Psi_q \Sigma\right]\text{Tr}[\Psi_{u} \Sigma^\dagger] +   M_{d} \text{Tr}\left[ \bar\Psi_{q'} \Sigma\right]\text{Tr}[\Psi_{d} \Sigma^\dagger] \label{fermionL} \end{align}

\subsection{Composite Higgs limit: CP assignment in the fermion sector}
As we show in the Appendix, loops of fermions and gauge bosons will generate a Coleman Weinberg potential at one loop, which will be of the form~\cite{0902.1483}
\bea V(\kappa,h) = a_1 h^2 + \lambda h^4 + |\kappa|^2 \left( a_2 + a_3 h^2 + a_4 |\kappa|^2 \right)  \,\,\,\,\,\,\,\text{ where }  \,\,\,\,\,\,\, \kappa = \sqrt{f^2-\eta^2 - h^2} + i \epsilon_t \eta \label{CHpot1}
\eea
where $a_i$ are dimensionful constants dependent on the form factors of the UV theory as given in the Appendix. Here $\epsilon_t$ is the parameter that defines the embedding of the up-type fermion in the global symmetry and determines the mass and CP assignment of $\eta$, as we demonstrated above. It is easy to see that the scenario in which $\epsilon_t$ is real is distinctly different from the case in which it can be complex.
For $\epsilon_t \in \mathbb{R} $, we find that $\eta$ behaves like a pseudoscalar ($c_{odd} \neq 0 $ and $c_{even} = 0$ in \eqref{CPferm}), and we can expand \eqref{CHpot1} to obtain the following CP and custodially symmetric potential: 
\bea  V(\eta,h) = m_h^2 h^2 + \lambda_h h^4 + m_\eta^2 \eta^2 + \lambda_\eta \eta^4 + c_4  \eta^2 h^2
\eea
Here, in terms of the parameters above we have defined 
\begin{subequations}
\begin{align} m_h^2 &= (a_1 + a_3 - a_2 - a_4) \\ \lambda_h &= (\lambda - a_3 + a_4) \\ m_\eta^2 &=(1-\epsilon_t^2) (-a_2 -a_4) \\ \lambda_\eta &= (1-\epsilon_t^2)^2 a_4 \\ c_4 &= (1-\epsilon_t^2) (-a_3 + 2 a_4) \end{align}
\end{subequations}
And as announced the trilinear term is absent. 
\iffalse 
It is seen that there will be a trilinear coupling in this scalar potential only if we expand $\eta$ around the mean field in the Hartree approximation ($\eta \rightarrow \bar\eta  + \tilde\eta$), 
\bea
V(\eta,h)  \ni c_4\bar\eta \tilde\eta  h^2
\eea
In this same Hartree approximation there will also be a momentum dependent trilinear Higgs coupling to $\eta$, by integrating by parts \eqref{eq:Lkin}
\bea 
\mathcal{L}_{kin} \ni - \frac{f^2}{2} h^2\bar\eta  \partial^2 \tilde\eta
\eea
As the kinetic Lagrangian is invariant under $(Z_2)^2$ (under both $h \rightarrow - h$, $\eta \rightarrow - \eta$ and under exchange of $\eta$ and $h$), the momentum dependence of this coupling will be a function of the total incoming four momentum $s = p_1 + p_2$.  
\fi
If we allow for complex coupling to fermions, 
\bea \epsilon_t = \epsilon_t^{RE} + i \epsilon_t^{IM} \eea
where $\epsilon_t^{IM} \neq 0$, we will find $\eta$ has $c_{even}\neq 0 $ in \eqref{CPferm}.\footnote{In the boundary case $\epsilon_t^{RE}=0, \epsilon_t^{IM} \neq 0$ $\eta$ behaves like a scalar. } 
%%%%
In this case the scalar potential will include a trilinear interaction and a tadpole for $\eta$, both of which multiply $\epsilon_t^{IM}$, 
\bea V = c_{tad} \eta + m_\eta^2 \eta^2  +\tilde{c}_\eta \eta^3 + \lambda_\eta \eta^4 +m_h^2 h^2  + \lambda_h h^4  + c_3 \eta  h^2  + c_4 \eta^2 h^2  \eea
where 
\begin{subequations}
\begin{align} \tilde{c}_\eta &= 4 a_4 \epsilon_t^{IM} \left(1-(\epsilon_t^{RE})^2\right) \sqrt{f^2-\eta^2-h^2} \\ \label{c3epsilon} c_3 &= (4 a_4- 2 a_3) (\epsilon^{IM}) \sqrt{f^2-\eta^2-h^2} \\ c_4 &= (a_3-2 a_4) (\epsilon^{RE})^2-4 a_4 (\epsilon^{IM})^2+2 a_4-a_3 \end{align}
\end{subequations}
and the other coefficients remain as above. The tadpole and trilinear interaction term violate CP for $\epsilon_t^{RE} \neq 0$.
%%%%
We may shift away the tadpole $c_{tad} \eta$ by an appropriate vacuum expectation value $v_\eta$, which solves,
\bea c_{tad} + 2 m_\eta^2 v_\eta + 3 \tilde{c}_\eta v_\eta^2 + 4 \lambda_\eta v_\eta^3 = 0 \eea
this will also shift the parameters,
\begin{subequations}
\begin{align} m_\eta^2 &\rightarrow  m_\eta^2 + 3 \tilde{c}_\eta v_\eta + 6 \lambda_\eta v_\eta^2 \\ 
\tilde{c}_\eta &\rightarrow \tilde{c}_\eta + 4 v_\eta \lambda_\eta \\ 
m_h^2 &\rightarrow m_h^2 + c_3 v_\eta + c_4 v_\eta^2 \\
c_3 &\rightarrow c_3 + 2 c_4 v_\eta  \end{align}
\end{subequations}
In terms of the shifted parameters the potential becomes
\bea \boxed{V =  m_\eta^2 \eta^2  +\tilde{c}_\eta \eta^3 + \lambda_\eta \eta^4 +m_h^2 h^2  + \lambda_h h^4  + c_3 \eta  h^2  + c_4 \eta^2 h^2}  \eea
This potential has the required form to be a suitable candidate for inflation followed by perturbative reheating.

\subsection{Spontaneously broken CP by the inflaton ($\alpha \neq 0$)}
For the Composite Higgs vacuum discussed above $\alpha = 0$ and CP is unbroken by the vacuum. Here we relax this constraint we introduce CP breaking in the model to 
\bea 0 < \alpha \leq 1/2 \pi \eea
For $\alpha = 1/2 \pi$ both fields have a quadratic term and do not interact. 
For the open interval, $0 < \alpha < 1/2 \pi$, we indeed find the same potential as at the end of the previous sector, to fourth order in the fields:
\bea V(\eta, h) =  m_\eta^2 \eta^2  +\tilde{c}_\eta \eta^3 + \lambda_\eta \eta^4 +m_h^2 h^2  + \lambda_h h^4  + c_3 \eta  h^2  + c_4 \eta^2 h^2\,. \label{eq:CPpot} \eea
The coefficients are in general nonzero, except for at $\alpha = 1/4 \pi$. We refer the reader to the Appendix for a discussion, and an example computation. Importantly, in these vacua we are not required to introduce explicit CP breaking by a complex fermion representation to get the $\eta$-odd terms as we were for $\alpha = 0$, that is, we may have either $  \in \mathbb{R}$ or $  \in \mathbb{C}$. 

In these vacua the $\eta$ field couples directly to fermions as 
\bea (\eta\, \bar{u}_R \, \slashed{p}\, u_R) \in \mathcal{L},\eea 
an effect proportional to $(1-   ^2)$. Indeed, is seen that the odd powers of $\eta$ in the potential (which includes the trilinear coupling) are multiplied by $(1-  ^2 )$ and $(b_1 - b_2\,  ^2) $ for some constants $b_i$ (from the linear and the second order expansion of the logarithm respectively). This combination plays the role that $\epsilon_t^{IM}$ played in the previous section, as an order parameter of CP breaking.

As expected from periodicity, the two quadrants $0 < \alpha < 1/2 \pi$ and $1/2 \pi < \alpha < \pi$ are equivalent, modulo a redefinition of the fields:\footnote{Because of custodial symmetry, which shows up here as a $Z_2$ symmetry for $h$, $h \rightarrow - h$ is a symmetry over the whole range. The latter substitution is therefore made for free.}
\bea \eta \rightarrow - \eta  \,\,\,\,\,\,\,\text{ and }  \,\,\,\,\,\,\,h \rightarrow - h \eea
We demonstrate this explicitly in the appendix.

We will finish this section with a comment on the appearance domain walls~\cite{Rubakov:1983bb}. As we introduced the possibility of breaking CP spontaneously, one may be worried that these will be present, and become energetically important. However, if the vaccuum breaks CP spontaneously, it does it at the scale of symmetry breaking $f$. But, as we will see in the next section, we expect inflation to occur below this scale, $\Lambda_{inf} < f$, hence the domain walls will be diluted during inflation.

\section{Inflation}
\iffalse
Setting the field $h$ to zero during inflation, we obtain the single field inflaton potential 
$$ V(\eta) =   \alpha \eta^2 + \beta \eta^4 + \frac{\alpha^2}{4 \beta}$$ 
gives inflation ($N=60$) with acceptable $n_s$ (and $r\approx 10^{-4}$) for $\beta/\alpha \approx - 2 \times 10^{-3}$
(I imposed  $\Lambda_c = \frac{\alpha^2}{4 \beta}$ such that $V(\eta_{min})=0$; this is necessary for polynomial inflation with a wrong sign mass term).
\fi

\begin{figure}[ht] 
\centering
  \includegraphics[width= 300pt]{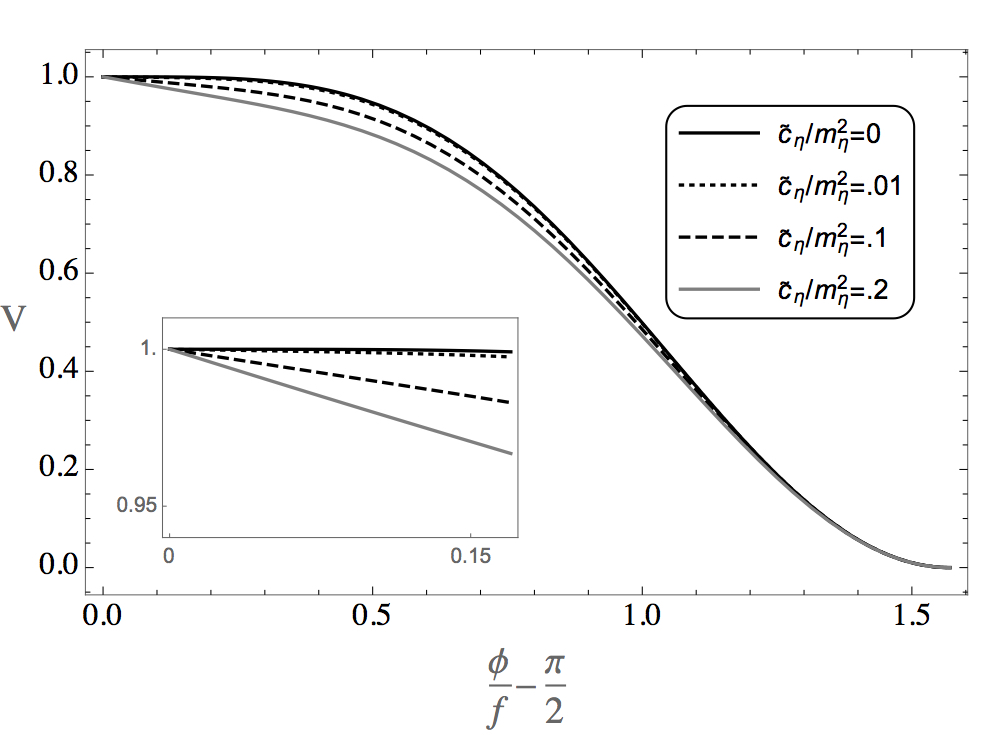}
  \caption{{\it Form of the potential:}  Here  $ \lambda_\eta f^2 \gtrsim - 1/2 m_\eta^2$. }\label{form}
\end{figure}

In this section we study inflation due to the field $\eta$. As the scale of inflation will turn out to be much larger than the electroweak scale, the Higgs field would be stabilized at the minimum of its potential during inflation, and so we set $h=0$. Hence, we neglect the dynamics of the Higgs field during inflation, and the model is effectively single field. We can canonically normalise the inflationary sector via the field redefinition
\bea \phi = f \arcsin\,( \eta /f)\,, \eea
such that the scalar potential becomes, in the unbroken CP limit,
\bea V_{CP}(\phi) = m_\eta^2 f^ 2 \sin( \phi/f )^2 + \lambda_\eta f^4 \sin (\phi/f)^4  \,.
   \eea
This is equivalent to the Goldstone Inflation \cite{Croon:2015fza} potential
\bea V(\phi) = \Lambda^4 \left(   \sin^2(\phi/f) - \tilde\beta \sin^4(\phi/f)\right)\,,\label{eq:infpot} \eea
if we identify 
$$\lambda_\eta f^4 = - \tilde\beta \Lambda^4  \,\,\,\,\,\,\,\text{ and }  \,\,\,\,\,\,\, m_\eta^2 f^2 =  \Lambda^4 . \label{relmass}$$ 
In figure \ref{form} we show a plot of the form of the potential, for the moment with $\tilde{c}_\eta /m_\eta^2 =0$.
This model would lead to inflation with $f< M_p$ (where $M_p$ is the reduced planck mass) and spectral index within the bounds allowed by Planck (at $2\sigma$) \cite{Ade:2015lrj}, 
$$ n_s = \left[.948 - .982 \right] \,\,\,\,\,\,\,\text{ for }  \,\,\,\,\,\,\, \tilde\beta \lesssim 1/2  \,\,\,\,\rightarrow \,\,\,\, \lambda_\eta f^2 \gtrsim - 1/2 m_\eta^2$$
As in Goldstone Inflation, the sensitivity to the exact value of $\tilde\beta$ that predicts the right spectral index is a function of $(f/M_p)^2$:
\bea \label{intervalbeta} 4  \times 10^{-4} \left(\frac{f}{M_p}\right)^2 < \delta \tilde\beta < 3 \times 10^{-3} \left(\frac{f}{M_p}\right)^2 \,\,\,\,\,\,\,\text{ where }  \,\,\,\,\,\,\, \delta\tilde\beta = 1/2 - \tilde\beta \eea
As in \cite{Croon:2015fza}, this feeds into the amount of tuning needed in the model, which we will discuss below. 

Likewise, the model has the initial condition for the start of slow roll as a function of $(f/M_p)^2$,
\bea \phi_i - 1/2 \pi f= (0.020 - 0.025 )\left( \frac{f}{M_p}\right)^2 M_p. \eea
As in all models of Goldstone Inflation, the tensor to scalar ratio will also be subject to fine tuning, but its value is generically very small:
\bea\label{rGI} r \approx 10^{-6} (f/M_p)^4\,.\eea 
A measurement of CMB tensor modes would fix the symmetry breaking scale $f$ (as well as the scale of inflation, as usual) in our model. 

\begin{figure}[ht] \label{Nsrplot}
\centering
  \includegraphics[width= 300pt]{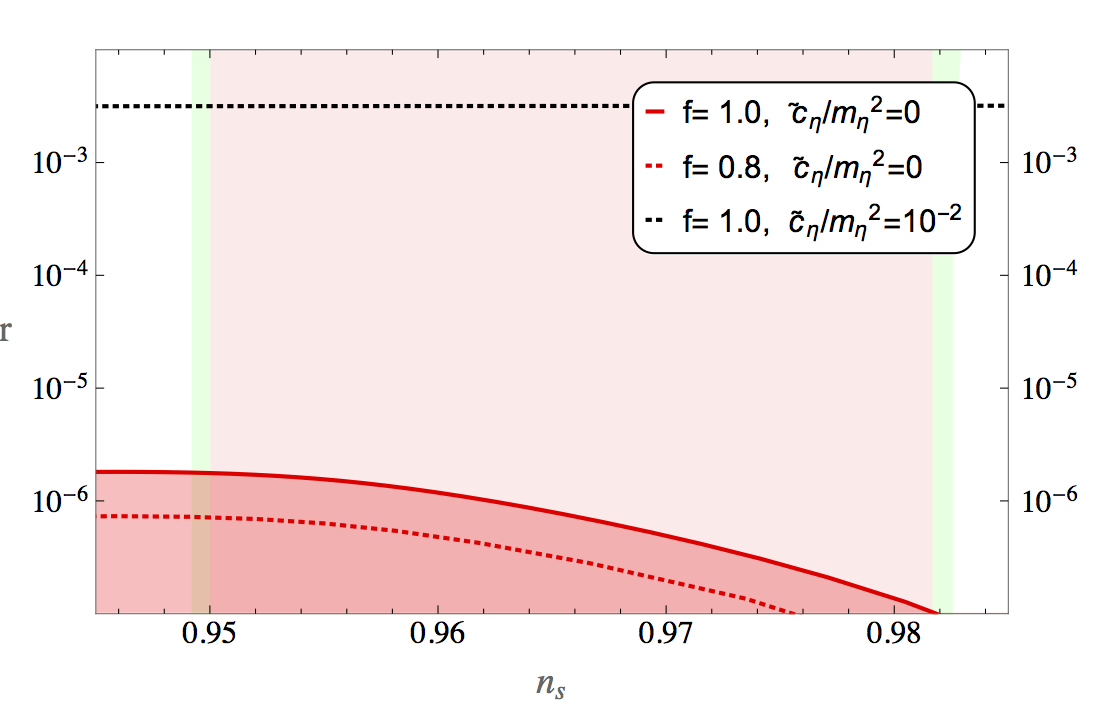}
  \caption{{\it Model predictions in the $n_s$-$r$ plane:} Planck 2015 $2 \sigma$ bounds \cite{Ade:2015lrj}. For convenience, we have set $M_p = 1$ here. In green: the TT spectrum and polarisation data at low-$\ell$ (lowP); in pink the combined spectra TT, TE, EE +lowP.}
\end{figure}

In the CP breaking fermion implementation described above there is an additional term  
\iffalse $$\epsilon_t^{IM} \sin (\phi/f)  \sqrt{1-\sin^2(\phi/f)} = \epsilon_t^{IM} \sin(\phi/f) \cos (\phi/f)$$ \fi 
\bea V_{\cancel{CP}}(\phi) =\tilde{c}_\eta \sin^3(\phi/f) \sqrt{1-\sin^2(\phi/f)} =\tilde{c}_\eta \sin^3(\phi/f) \cos (\phi/f)\,. \eea 
This term imposes modulations on the potential with period $ \pi f$, as seen from Fig.~\ref{form}. Increasing the CP breaking in the model corresponds to increasing the value of the tensor to scalar ratio $r$. The bound $r < 0.1$ gives 
\bea \tilde{c}_\eta \leq \mathcal{O}(10^{-1})\, m_\eta^2  f^2\,.\eea
The effect of the CP breaking term is illustrated for an order of magnitude below this bound in Fig.~\ref{Nsrplot}.

The scale of inflation is related to the amplitude of the scalar power spectrum, as measured by Planck \cite{Ade:2015lrj}, 
\bea A_s = \frac{\Lambda^4 }{24 \pi^2 M_p^4 \epsilon } = \frac{e^{3.089}}{10^{10}}\eea
where $\epsilon$ is the first slow roll parameter. For our case (Eq.~(\eqref{rGI}), where $r = 16 \epsilon $ in the slow roll approximation) this implies
\bea \Lambda \approx 10^{15} \left(\frac{f}{M_p} \right) \text{GeV}\,.\label{eq:infscale}\eea 
Interestingly, we can see from this relation that the onset of inflation is related to the scale of the symmetry breaking: $\Lambda \sim 10^{-3} f$. That is, fitting to the CMB data implies a mass gap of roughly three orders of magnitude between the two scales. 

\subsection{Tuning}
Following convention, tuning can be expressed numerically using the Barbieri-Giudice \cite{Barbieri:1987fn} parametrization as follows
\begin{equation}\label{Delta} \Delta = \left|  \frac{\partial\, \log n_s}{\partial \log \tilde{\beta}} \right| = \left| \frac{\tilde{\beta}}{n_s} \frac{\partial\, n_s}{\partial \tilde{\beta}} \right| \approx \left[ 8.1 - 8.5 \right] \left( \frac{f}{M_p}\right)^{-2} \end{equation}
See Fig. \ref{TuningPlot} below. It is seen that the parameters are sensitive to the square of the ratio of scales.
\begin{figure}[ht] 
\centering
  \includegraphics[width= 300pt]{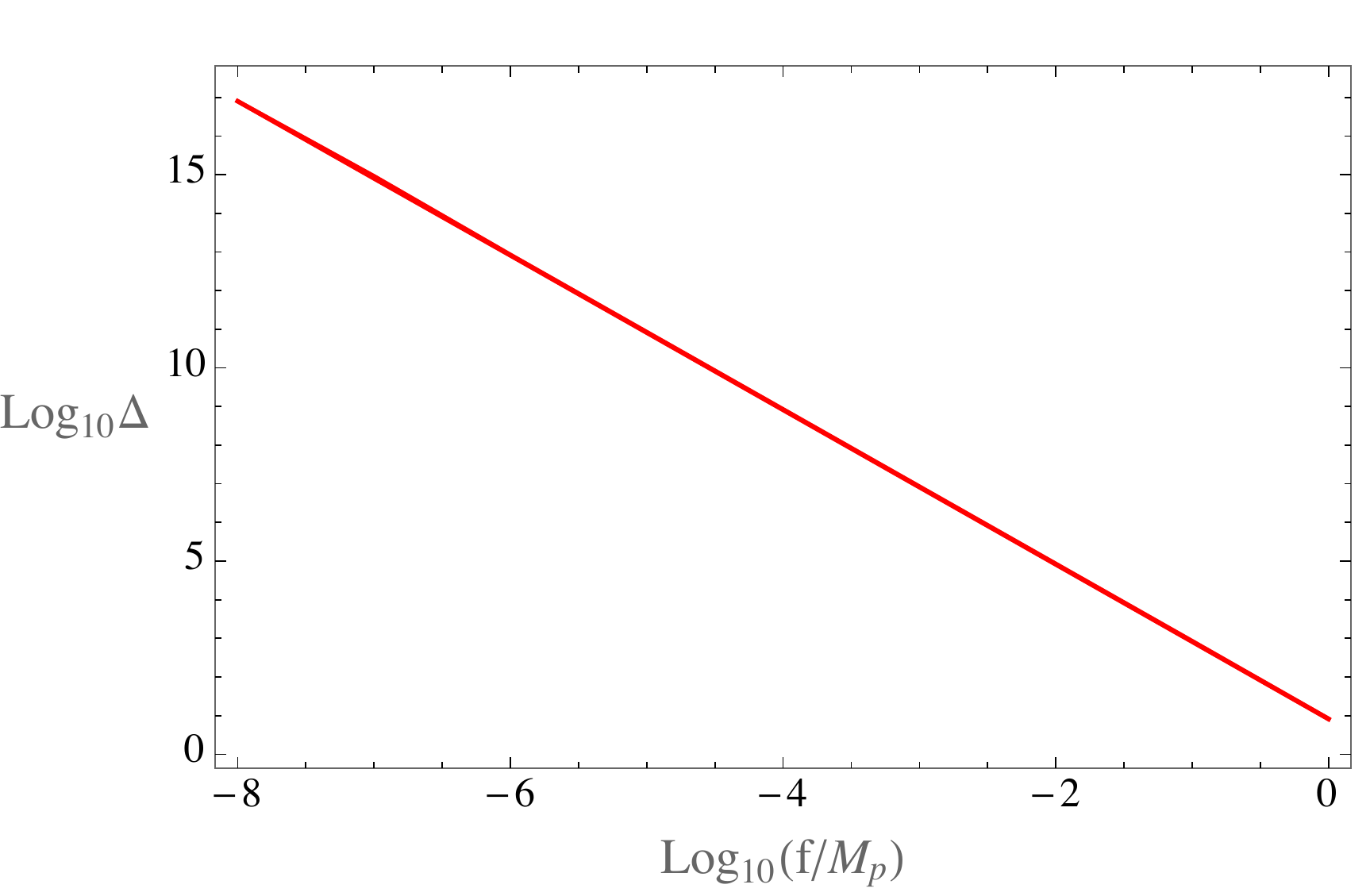}
  \caption{{\it Fine-tuning:} numerically defined as in \eqref{Delta}.}\label{TuningPlot}
\end{figure}

However, the relation $\tilde\beta \approx .5 $ can be seen as a consequence of a symmetry in the sector responsible for the breaking of the global symmetry $SO(6)/SO(5)$. This would agree with naturalness in the 't Hooft interpretation. In this case the fact that the small deviation $\delta \tilde\beta$ is sensitive to the relation of the scales $f$ and $M_p$ implies that a symmetry in the sector is broken at the same time as  $SO(6)/SO(5)$. In \cite{Croon:2015fza} we related this symmetry to the spectrum of resonances in the composite sector.

When we identify the other scalar resonance with the Higgs, we introduce a second source of tuning, between the electroweak scale $v$ and the symmetry breaking scale $f$. This source of tuning coincides with the tuning in the Minimal and the Next to Minimal Composite Higgs model, and is a function of $(v/f)^2$, see for instance \cite{Agashe:2004rs}. As this is a tuning of the parameters in the Higgs potential, which are independent combinations of the input parameters (the form factors, vacuum angles, and fermion representation), this tuning is independent and additive. The Barbieri-Giudice function will then take the form $\Delta_{\text{total}} = c_1 \left(M_p/f\right)^{2} + c_2 \left(f/v\right)^{2}$, where $c_1$ and $c_2$ are $\mathcal{O}(1)$ constants. 
This suggests that the Barbieri-Giudice function is minimized for $\Delta_{\text{total}} (f^2= \sqrt[4]{c_1/c_2}\, M_p v) \sim 10^{16}$, which is a large, but technically natural fine-tuning.

%%%%%%%%%%%%%%%%%%%%%%%%%%%%%%%%%%%%%%%%%%%%
\section{Reheating}\label{sec:reheating}
%%%%%%%%%%%%%%%%%%%%%%%%%%%%%%%%%%%%%%%%%%%%

At the end of inflation, the inflation field approaches, overshoots and begins to oscillate about the minimum of its potential.  At this stage, the universe is completely dominated by the zero--mode of the oscillating inflaton field $\langle\phi(t)\rangle$.  Interactions with the higgs field, which we have so far neglected, lead to dissipation which drains energy from $\langle\phi(t)\rangle$, and excites relativistic higgs particles. We refer to these collective processes as reheating (see e.g.,~\cite{Kofman:1997yn,Allahverdi:2010xz} for reviews).  The calculation that we present below section is semi--classical: we treat the inflaton condensate as a classical source in the mode equations for the quantum fluctuations of the higgs field. This treatment neglects many of the complicated processes which are present during the reheating phase, such as thermal corrections, re--scatterings of the produced higgs particles on the inflaton condensate, and the thermalisation process. As we discuss at the end of this section, these effects can in general modify the rate of decay of the condensate. Our approach does however provide an estimate for the perturbative decay rate of $\langle\phi(t)\rangle$ into higgs particles, and allows us to estimate the reheating temperature $T_R$.

%----------------------------------------------------------------------
\subsection{Equations of Motion}

\begin{figure}[h!]
\begin{center}
\includegraphics[width=17cm]{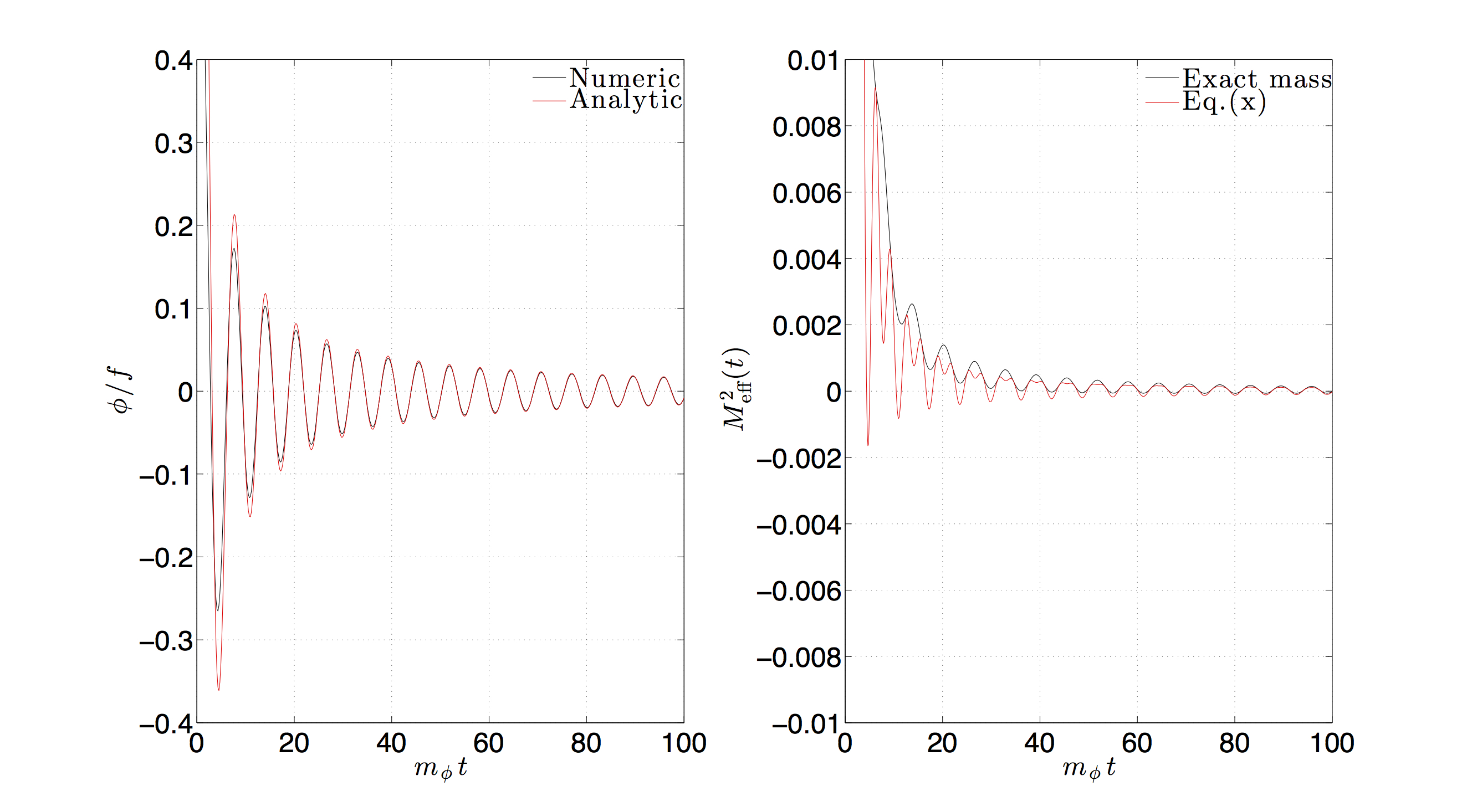}
\caption{\emph{Left panel}: Comparison between the exact numerical solution of Eq.~(\ref{eq:phiEoM}) and the approximate analytic solution Eq.~(\ref{eq:phisol}). \emph{Right panel}: Comparison between the exact `mass' (the coefficient of the term linear in $\chi$ of Eq.~(\ref{eq:chimastereq})) and $M^2_{\rm eff}(t)$ as defined in Eq.~(\ref{eq:M2eff}).}
\label{fig:phi-sol-M2eff_t}
\end{center}
\end{figure}

To begin, we study the classical inflaton background. As a first approximation, we neglect interactions with the higgs field and set $h=0$. As before, the inflaton sector can be canonically normalised through the field redefinition $\eta(t)=f\sn(\phi(t)/f)$. We neglect excitations of the inflaton field, $\delta\phi$, and so for simplicity label the zero--mode $\phi(t)\equiv\langle\phi(t)\rangle$ which obeys the usual Klein Gordon equation:
\beq
\ddot\phi + 3H\dot\phi + \frac{\de V}{\de \phi}\Big|_{h=0}=0\,,
\label{eq:phiEoM}
\eeq
where the potential is given by Eq.~(\ref{eq:infpot}). After inflation, the inflaton field approaches, overshoots and begins to oscillate about its minimum. This region of the potential, where $\phi/f\ll1$, is essentially quadratic:
\beq
V_{h=0}(\phi) \approx \frac12\mphi^2\phi^2\,, \qquad
\mphi^2 \equiv 2m_\eta^2 \approx 
2\times10^{-14}\left( \frac{f}{\Mp}\right)^2\Mp^2  \,,
\label{eq:mphidef}
\eeq
where we have used the Planck constraint on the amplitude of scalar power spectrum (Eq.~(\ref{eq:infscale})) to determine the mass $\mphi$ in terms of the scale $f$. To describe the oscillations, notice that Eq.~(\ref{eq:phiEoM}) can be written as
\beq
\frac{{\rm d}^2}{{\rm d}t^2}(a^{3/2}\phi) + \left[ \mphi^2 - \left( \frac94H^2+\frac32\dot H \right)\right](a^{3/2}\phi)=0\,.
\label{eq:phiEoM2}
\eeq
At the onset of oscillation, $\mphi^2\gg H^2,\dot H$ and under this condition, Eq.~(\ref{eq:phiEoM2}) has the damped sinusoidal solution:
\beq
\phi(t)=\frac{\Phi_0}{a^{3/2}(t)}\,{\rm sin}\,\left(\mphi t + \vartheta \right) \,, \qquad 
\Phi_0 \approx 0.6 \left(\frac{f}{\Mp}\right) \Mp\,.
\label{eq:phisol}
\eeq
The numerical value for the initial amplitude, $\Phi_0$, was obtained by matching the above solution with an exact numerical integration of Eq.(\ref{eq:phiEoM}) -- see the left hand panel of Fig.~\ref{fig:phi-sol-M2eff_t} for illustration. Subscript zero denotes evaluation at the onset of oscillations (start of reheating), and we set $a_0=1$. The scale factor, averaged over many oscillations, grows as $a(t)\sim t^{2/3}$, while the energy density of the field decreases as:
\beq
\rho_\phi(t)=\frac12\dot\phi^2(t) + \frac12\mphi^2\phi^2(t) \simeq \frac{\mphi^2\Phi_0^2}{2a^3}\,.
\eeq
We see that the vacuum energy of the inflaton field exists as spatially coherent oscillations, which can be interpreted as a condensate of non--relativistic zero--momentum $\phi$--particles. The amplitude of the oscillations decay due to the Hubble expansion and also due production of higgs particles. We can obtain an estimate for this particle production rate by considering propagation of higgs fluctuations, $h_k$, in the background of the classical inflaton condensate.\\

We begin by canonically normalising the higgs kinetic sector (given by Eq.~(\ref{eq:Lkin})) by performing the following field redefinition:
\beq
\de_\mu\chi(x) = \sqrt{\frac{f^2-\eta^2(t)}{f^2-\eta^2(t)-h^2(x)}}\de_\mu h(x)\,,
\eeq
such that
\beq
h(x) = f \cn(\phi(t)/f)\,\sn\hat\chi(x)\,,  \qquad 
\hat\chi(x) \equiv \frac{\chi(x)}{f\cn^2(\phi(t)/f)}\,.
\eeq
We will henceforth drop the space--time labels and write $\chi=\chi(x)$, $\phi=\phi(t)$: it is to be understood that the higgs is inhomogeneous, whilst the inflaton condensate is homogeneous, and described by Eq.~(\ref{eq:phisol}). Under these field redefinitions we obtain:
\beq
\mathcal{L}=-\frac12\de_\mu\chi\de^\mu\chi  
- \frac12\left[ 1 + \sn^2(\phi/f)\tn^2\hat\chi \right]\de_\mu\phi\de^\mu\phi
-\left[ \sn(\phi/f)\,\tn\hat\chi\right] \de_\mu\chi\de^\mu\phi - V(\phi,\chi)\,,
\label{eq:higgsL}
\eeq
where the potential is given by Eq.~(\ref{eq:CPpot}). The canonically normalised higgs equation of motion is obtained by varying the action with respect to $\chi$:
\beq
\ddot\chi - \frac{\nabla^2}{a^2}\chi + 3H\dot\chi = - \frac{\de V(\phi,\chi)}{\de \chi} + 
{\rm sin}\,(\phi/f)\,{\rm tan}\,\hat\chi \,\frac{\de V(\phi)}{\de\phi}\Big|_{h=0}
-\frac{\dot\phi^2}{f^2}\,K(\phi,\chi)\,,
\label{eq:chimastereq}
\eeq
where 
\bea
K(\phi,\chi)\equiv \frac{f\sn\hat\chi\,\cn^2\hat\chi\,\cn^4(\phi/f) +2\chi\cn\hat\chi\,\sn^2(\phi/f) - f\sn\hat\chi\,\cn(\phi/f) + f\sn\hat\chi\,\cn^3(\phi/f)}{\cn^3(\phi/f)\,\cn^3\hat\chi}\,.
\label{eq:Kdef}
\eea
In deriving Eq.~(\ref{eq:chimastereq}), we have used Eq.~(\ref{eq:phiEoM}) to eliminate $\ddot\phi$ which arises from the variation of the action. The task at hand is to solve Eq.~(\ref{eq:chimastereq}) given the inflaton background Eq.~(\ref{eq:phisol}). This is made tractable by expanding the RHS of Eq.(\ref{eq:chimastereq}) about $\phi/f=0$, and about $\chi/f=0$:
\beq
\ddot\chi - \frac{\nabla^2}{a^2}\chi + 3H\dot\chi \approx 
-\left[  \mchi^2 + \tri\phi + \qtc\phi^2 + \frac{\dot\phi^2}{f^2}  \right]\chi + \cdots\,,
%-\left[    \lambda_\chi + y\phi(t) + \xi^2\phi^2(t) + \frac{\dot\phi^2}{f^4}  \right]\chi^3\,,
\label{eq:higgsEoMfinal}
\eeq
where we have defined
\bea
m_\chi^2 &\equiv& 2m_h^2 \,, \qquad
 \tri \equiv 2c_3\,, \qquad 
\qtc \equiv  2\left[m^2_h/f^2-m^2_\eta/f^2+c_4 \right]\,.
%\lambda_\chi &\equiv& \frac{4}{f^4}\left[ \lambda_h - \frac{m_h^2}{3} \right]\,, \qquad
 %y \equiv -\frac{4c_3}{3f^5}\,, \qquad 
%\xi^2 \equiv  -\frac{2}{f^6}\left[ 2m_h^2 - 4\lambda_h + \frac{m^2_\eta}{3} +\frac{2c_4}{3}  \right]\,.
\eea
The expansion in $\phi/f$ is permitted since the amplitude of the inflaton oscillations are small with respect to the scale $f$: $\Phi_0/a^{3/2}(t)\sim0.6f/a^{3/2}(t)$. The expansion in $\chi/f$ is permitted since we assume that the higgs field is stabilised at the minimum of its potential throughout inflation, $\langle \chi({\bf x},t)\rangle=0$. Furthermore we consider perturbative reheating only: we restrict ourselves to regions of parameter space where the coupling constants $\tri$ and $\qtc$ are small enough such that resonant enhancement of higgs modes is not possible. This ensures that $\chi\ll f$ throughout reheating. We will discuss the conditions for perturbative reheating shortly. Notice that inflaton mass, $m_\eta^2$, and the higgs mass, $m_h^2$, enter the definition of the coupling $\qtc$: their presence may be traced back to canonical normalisation of the higgs kinetic term.  \\

For the analysis of Eq.~(\ref{eq:higgsEoMfinal}) it is convenient to define a co--moving field
\beq
\mu_k(\tau)\equiv a(\tau)\chi_k(\tau)\,,
\eeq
and to work in conformal time, which is related to cosmic time by an integral over the scale factor:
\beq
t(\tau)=\int^{\tau}_{\tau_0} {\rm d}\tau' a(\tau')\,.
\eeq
According to standard arguments, we may decompose this field into creation and annihilation operators:
\beq
\mu(\tau,{\bf x}) = \int \frac{{\rm d}^3k}{(2\pi)^{3/2}} \left[ a_{\bf k}\mu_k(\tau) + a_{\bf -k}^\dagger\mu^*_k(\tau)  \right] e^{i{\bf k\cdot x}}\,,
\eeq
where the mode functions obey
\beq
\mu''_k(\tau) + \omega^2_k(\tau)\mu_k(\tau)=0\,,
\label{eq:modeEoM}
\eeq
and where a prime denotes differentiation with respect to conformal time. The time dependent frequency is given by 
\beq
 \omega_k^2(\tau)\equiv k^2+a^2M^2_{\rm eff}(\tau) - \frac{a^{\p\p}}{a} \,, \qquad
\frac{a^{\p\p}}{a}  = \frac{a^2}{6\Mp^2}\left( \rho_\phi - 3P_\phi \right)\,,
\label{eq:omegak}
\eeq
where $P_\phi\simeq0$ is the pressure of the field, and we have defined the effective mass:
\beq
M^2_{\rm eff}(t) \equiv  \mchi^2 + \frac{\tri\Phi_0}{a^{3/2}(t)}\,\sn(\mphi t+\vartheta) + \frac{\qtc\Phi^2_0}{a^3(t)}\,\sn^2(\mphi t+\vartheta) +
\frac{\Phi^2_0\mphi^2}{f^2a^3(t)}\,\cn^2(\mphi t+\vartheta) \,.
\label{eq:M2eff}
\eeq
The final term on the RHS of $M^2_{\rm eff}(t)$ is the leading contribution from $\dot\phi^2/f^2$: we have neglected terms which decay faster than $a^{-3}$. In the right panel of Fig.~\ref{fig:phi-sol-M2eff_t}, we plot the effective mass against the coefficient of the term linear in $\chi$ of Eq.~(\ref{eq:chimastereq}), which demonstrates the accuracy of this expansion. Equations of the type (\ref{eq:modeEoM}), with time dependent mass (\ref{eq:M2eff}) have been extensively studied in the context of (p)reheating after inflation. For certain regions of $\{\tri,g^2,\Phi_0\}$ parameter space, the mode functions experience exponential growth as parametric instability develops, a phenomenon known as parametric resonance~\cite{Traschen:1990sw,Shtanov:1994ce,Kofman:1997yn,Braden:2010wd}. To be specific, when any one of the three terms in $M^2_{\rm eff}(t)$ is dominant, the oscillator equation~(\ref{eq:modeEoM}) may written 
\beq
\frac{{\rm d}^2\mu_k}{{\rm d}z^2}+\left[ A_k -2q_i\,\cn(2z) \right]\mu_k=0\,,
\label{eq:Mathieu}
\eeq
\beq
q_0\equiv \frac{\Phi_0^2}{4f^2a^3}\,, \qquad
q_3\equiv \frac{\tri\Phi_0}{\mphi^2a^{3/2}}\,, \qquad
q_4\equiv \frac{\qtc\Phi_0^2}{4\mphi^2a^3}\,, \qquad
A_k\equiv \frac{k^2+\mchi^2}{\mphi^2a^2}+2q_{(0,4)}\,,
\eeq
following a time redefinition of the form $z\equiv \mphi t+{\rm const}$. Here we have ignored terms proportional to $H/\mphi$ (recall that $H\ll\mphi$ during reheating). Eq.~(\ref{eq:Mathieu}) is known as the Mathieu equation, which is known to possess instability bands for certain values of $A_k$ and $q_i$. For $q_i\gg1$, a large region of parameter space is unstable and broad parametric resonance can develop. Throughout this paper we restrict ourselves to regions of parameter space where $q_i\ll1$, such that non--perturbative preheating processes are negligible. With $\Phi_0\approx 0.6f$, we find $q_0=0.09$, and so parametric instability cannot be triggered by this term. Meanwhile, $q_{3,4}\ll1$ requires:
\beq
\tri \ll \frac{\mphi^2}{\Phi_0}\,, \qquad 
\qtc \ll \left(  \frac{\mphi}{\Phi_0} \right)^2\,,
\label{eq:pert}
\eeq
or, in terms of the original parameters of the potential~(\ref{eq:CPpot}):
\beq
c_3\ll m_\eta^2/f\,, \qquad
m_h^2/f^2 + c_4\ll 10m_\eta^2/f^2\,. \label{orig:pert}
\eeq
This relation for the smallness of the CP breaking term $c_3$ in terms of the inflaton mass is consistent with the similar relation for $c_\eta$ found in the previous section. Likewise, the constraint on $c_4$ is consistent with our expectations from the computation of the potential, as can be verified with the appendix. We always ensure that the above bounds are respected, and do not consider parametric resonance in this paper.\\

If we regard the inflaton condensate $\phi$ to be a collection of zero--momentum inflaton `particles', then the effective mass $M^2_{\rm eff}(t)$ has a physical interpretation in terms of Feynman diagrams:
\begin{figure}[h!]
\begin{center}
\includegraphics[width=8cm]{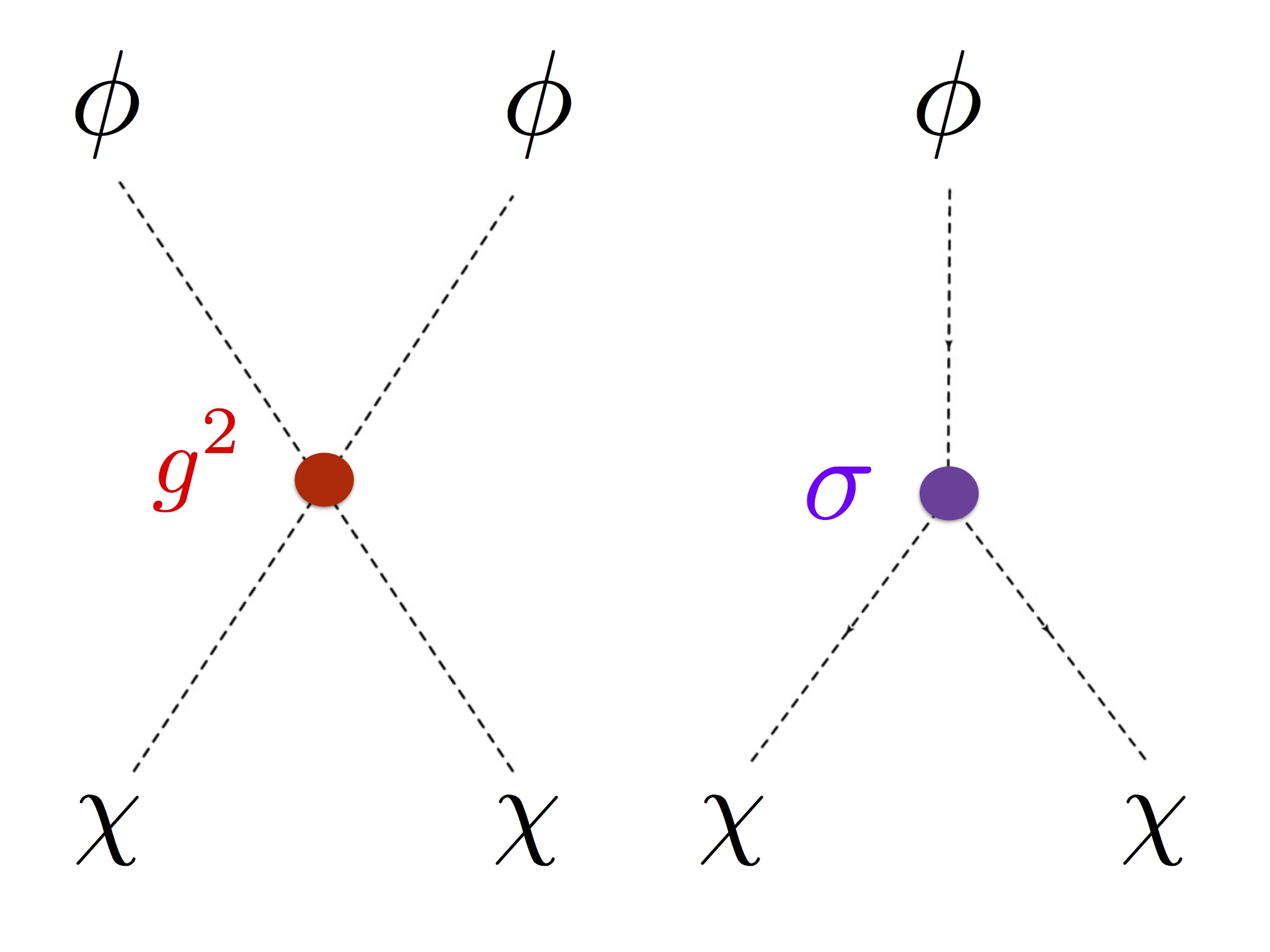}
\label{fig:vertices}
\end{center}
\end{figure}

These diagrams describe the three--leg, $-\frac12\tri\phi\chi^2$, and four--leg, $-\frac12\qtc\phi^2\chi^2$, interaction terms which reside in the canonically normalised Lagrangian -- Eq.~(\ref{eq:higgsL}). Since we have not quantised the inflaton, there are no $\phi$--propagators, which allows for tree--level diagrams only. These diagrams describe the perturbative decay of a single inflaton `particle' with mass $\mphi$ into two higgs particles of comoving momentum $k\sim a\mphi/2$, and the annihilation of a pair of $\phi$ `particles'  into pair of $\chi$ particles with comoving momentum $k\sim a\mphi$ respectively. We use the term inflaton `particle' rather loosely here, since what we are really describing is creation of higgs particles from a classical inflaton condensate. This diagrammatic representation does however offer intuition for the physical processes at work.

%----------------------------------------------------------------------
\subsection{Bogoliubov Calculation}\label{sec:Bogoliubov}

We wish to solve Eq.~(\ref{eq:modeEoM}) with frequency (\ref{eq:omegak}). Our calculation closely follows that of Ref.~\cite{Braden:2010wd}. First, we notice that since the inflaton condensate behaves like a collection of non--relativistic particles with zero pressure, $P_\phi\approx0$, and so we have $a''/a\approx2a^2H^2$. Therefore, for the modes $k^2\sim a^2\mphi^2$ which we expect to be produced, we can safely neglect $a''/a$, given that $H\ll\mphi$ during reheating. In the adiabatic representation, the solution to the mode equation Eq.~(\ref{eq:modeEoM}) may be written in the WKB form (see eg.~\cite{Kofman:1997yn,Braden:2010wd}):
\beq
\mu_k(\tau) = \frac{\alpha_k(\tau)}{\sqrt{2\omega_k(\tau)}} e^{-i\Psi_k(\tau)} + \frac{\beta_k(\tau)}{\sqrt{2\omega_k(\tau)}} e^{+i\Psi_k(\tau)}\,,
\label{eq:WKB}
\eeq
where the accumulated phase is given by
\beq
\Psi_k(\tau') \equiv \int^{\tau'}_{\tau_0}  {\rm d}\tau'' \omega_k(\tau'')\,.
\eeq
Eq.~(\ref{eq:WKB}) is a solution of Eq.~(\ref{eq:modeEoM}) provided that the Bogoliubov coefficients satisfy the following coupled equations:
\beq
\alpha_k'(\tau) = \beta_k(\tau)\frac{w_k'(\tau)}{2w_k(\tau)}e^{+2i\Psi_k(\tau) }\,, \qquad
\beta_k'(\tau) = \alpha_k(\tau)\frac{w_k'(\tau)}{2w_k(\tau)}e^{-2i\Psi_k(\tau) }\,,
\label{eq:alphabeta}
\eeq
which also implies that:
\beq
\mu'_k(\tau) =  -i\alpha_k(\tau)\sqrt{\frac{w_k(\tau)}{2}}e^{-i\Psi_k(\tau)} + i\beta_k(\tau)\sqrt{\frac{w_k(\tau)}{2}}e^{+i\Psi_k(\tau)}\,.
\eeq
The wronskian condition, $W[\mu_k(t),\mu_k^*(t)]=i$, demands that the Bogoliubov coefficients are normalised as $|\alpha_k(t)|^2-|\beta_k(t)|^2=1$. In this basis, the Hamiltonian of the $\chi$ field is instantaneously diagonalised. The single particle mode occupation number $n_k$, is defined as the energy of the mode, $\frac12|\mu_k'|^2+\frac12\omega^2_k|\mu_k|^2$, divided by the frequency of the mode:
\beq
n_k(\tau) = \frac{ |\mu'_k(\tau)|^2 +\omega^2_k(\tau)|\mu_k(\tau)|^2  }{2\omega_k(\tau)} -\frac12 = |\beta_k(\tau)|^2\,.
\eeq
The $-1/2$ corresponds to subtraction of the zero--point energy, and the last equality is obtained via substitution of the WKB solution~(\ref{eq:WKB}). In terms of the classical mode functions, creation of higgs particles occurs due to departure from the initial positive--frequency solution: the initial conditions therefore at $\tau=\tau_0$ (the start of reheating) are then $\alpha_k=1$, $\beta_k=0$, and so $n_k(\tau_0)=0$. Since we work in the perturbative regime specified by Eq.~(\ref{eq:pert}) the mode occupation numbers remain small, $|\beta_k(\tau)|^2\ll1$, and so we can iterate Eq.~(\ref{eq:alphabeta}) to obtain
\beq
\beta_k(\tau) \approx \int^{\tau}_{\tau_0} {\rm d}\tau' \, \frac{\omega'_k(\tau')}{2\omega_k(\tau')}\, e^{-2i\Psi_k(\tau')}\,.
\label{eq:betasol}
\eeq
In the perturbative regime we can approximate
\beq
\Psi_k(\tau') \approx k\int^{\tau'}_{\tau_0}  {\rm d}\tau'' \sqrt{1+\left(\frac{a(\tau'') \mchi}{k}\right)^2 }\,,
\eeq
whilst for the frequency we have
\beq
\frac{\omega'_k}{2\omega_k} \approx \frac{a^{3/2}(\tau')\Phi_0\mphi}{4k^2}
\left[ \frac{\tri+2\Phi_0(\qtc-\mphi^2/f^2)a^{-3/2}(\tau')\sn(\mphi t(\tau')+\vartheta)}{1+a^2(\tau')\mchi^2/k^2}\right]\,\cn(\mphi t(\tau')+\vartheta)\,,
\eeq
where we have neglected terms containing derivatives of the scale factor. Inserting these results into Eq.~(\ref{eq:betasol}) gives:
\bea
\beta_k(\tau) &=& \frac{\tri\Phi_0\mphi}{8k^2}\int^{\tau}_{\tau_0} \frac{{\rm d}\tau'\,a^{3/2}(\tau')}{1+a^2(\tau')\mchi^2/k^2} \left[  e^{+i\psi^-_{3,k}(\tau')} + e^{-i\psi^+_{3,k}(\tau')} \right] \nonumber \\
  &+&  \frac{(\qtc-\mphi^2/f^2)\Phi^2_0\mphi}{8ik^2}\int^{\tau}_{\tau_0} \frac{{\rm d}\tau'}{1+a^2(\tau')\mchi^2/k^2} \left[  e^{+i\psi^-_{4,k}(\tau')} - e^{-i\psi^+_{4,k}(\tau')} \right]\,,
\label{eq:betasol2}
\eea
where we have defined the phases
\beq
\psi^\pm_{3,k}(\tau) \equiv \pm2\Psi_k(\tau) + \mphi t(\tau) + \vartheta\,, \qquad
\psi^\pm_{4,k}(\tau) \equiv \pm2\Psi_k(\tau) + 2(\mphi t(\tau) + \vartheta)\,.
\eeq
As discussed in Ref.~\cite{Braden:2010wd}, (see also~\cite{Kofman:1997yn}), the integrals in Eq.~(\ref{eq:betasol2}) can be evaluated using the method of stationary phase: they are dominated near the instants $\tau_{3,k}$ and $\tau_{4,k}$ where
\bea
\frac{{\rm d}}{{\rm d} \tau} \psi^-_{3,k}(\tau) \Big|_{\tau_{3,k}} = 0\,, 
  & \Rightarrow &
k=\frac12\mphi a(\tau_{3,k})\sqrt{1-4\delta_M^2}\,, \nonumber \\
\frac{{\rm d}}{{\rm d} \tau} \psi^-_{4,k}(\tau) \Big|_{\tau_{4,k}} =0\,,  
  & \Rightarrow &
 k=\mphi a(\tau_{4,k})\sqrt{1-\delta_M^2}\,,
\label{eq:k-modes}
\eea
where we have defined $\delta_M \equiv \mchi/\mphi$. For the 3--leg interaction, the above result corresponds to the creation of pair of higgs particles with momentum $k\sim a\mphi/2$ from an inflaton with mass $\mphi$ at the instant $\tau_{3,k}$ of the resonance between the mode $k$ and the inflaton condensate.  A similar interpretation may be given for the 4--leg interaction. Upon performing the integrals, we find:
\bea
n_k(\tau) &=& \frac{\pi\tri^2\Phi_0^2\mphi}{32k^4}\left( 1-4\delta_M^2\right)  \frac{a^3(\tau_{3,k})}{a'(\tau_{3,k})} 
+ \frac{\pi (\qtc-\mphi^2/f^2)^4\Phi_0^2\mphi}{64k^4} \frac{\left( 1-\delta_M^2 \right) }{a'(\tau_{4,k})} \nonumber \\
&+& \frac{\pi\tri(\qtc-\mphi^2/f^2)\Phi_0^3\mphi}{32k^4}\sqrt{2\left( 1-4\delta_M^2\right)\left( 1-\delta_M^2\right)}\,\mathcal{I}(\tau_{3,k}\,\tau_{4,k})\,,
\eea
where we have defined
\beq
\mathcal{I}(\tau_{3,k}\,\tau_{4,k}) \equiv  \sqrt{\frac{a^3(\tau_{3,k})}{a'(\tau_{3,k})a'(\tau_{4,k})}}\,\sn\left[ \psi^-_{4,k}(\tau_{4,k}) - \psi^-_{3,k}(\tau_{3,k})\right]\,.
\eeq
As discussed in~\cite{Braden:2010wd}, the oscillatory term $\mathcal{I}(\tau_{3,k}\,\tau_{4,k})$ represents the interference between the two decay channels ($\phi\to\chi\chi$ and $\phi\phi\to\chi\chi$) of the inflaton. It is present because we have treated the inflaton as a classical oscillating source, and not an honest collection of particles.

%----------------------------------------------------------------------
\subsection{Boltzmann Equations}

Since $\mphi\gg\mchi$ the higgs particles are relativistic when produced. This means we can effectively treat them as a bath of radiation with $g_*$ number of degrees of freedom.  We define the co--moving energy density in the higgs field as
\bea
a^4\rho_\chi &\equiv& \int^\infty_0 \frac{{\rm d}^3k}{(2\pi)^3}\,\omega_k n_k  \nonumber \\
&=&\frac{\tri^2\Phi_0^2\mphi}{64\pi} \left( 1-4\delta_M^2 \right) \int^{\infty}_{0} \frac{{\rm d}k}{k^2}\sqrt{k^2+a^2(\tau)\mchi^2}\,\frac{a^3(\tau_{3,k})}{a'(\tau_{3,k})} \nonumber \\
&+&\frac{(\qtc-\mphi^2/f^2)^2\Phi_0^4\mphi}{128\pi} \left( 1-\delta_M^2 \right) \int^{\infty}_{0} \frac{{\rm d}k}{k^2} \sqrt{k^2+a^2(\tau)\mchi^2}\,\frac{1}{a'(\tau_{4,k})} \nonumber \\
&+& \frac{\tri(\qtc-\mphi^2/f^2)\Phi_0^3\mphi}{64\pi}\sqrt{2\left(1-4\delta_M^2\right)\left(1-\delta_M^2\right)}\int^{\infty}_{0} \frac{{\rm d}k}{k^2}\sqrt{k^2+a^2(\tau)\mchi^2}\,\mathcal{I}(\tau_{3,k}\,\tau_{4,k})\,. \nonumber \\
\label{eq:boltz1}
\eea
At first glance these integrals appear divergent. This however is not the case, as can be seen from the requirement that the higgs particles be produced perturbatively. Eq.~(\ref{eq:k-modes}) enforces:
\bea
\frac12\mphi a_0\sqrt{1-4\delta^2_M}<&k&<\frac12\mphi a(\tau)\sqrt{1-4\delta^2_M}\,, \qquad {\rm for}\,\,\,\, \phi\to\chi\chi  \label{eq:energycon3}\\
\mphi a_0\sqrt{1-\delta^2_M}<&k&<\mphi a(\tau)\sqrt{1-\delta^2_M}\,, 
\qquad {\rm for}\,\,\,\, \phi\phi\to\chi\chi\,.
\label{eq:energycon4}
\eea
Hence, the limits of the first and the third integrals on the RHS of Eq.~(\ref{eq:boltz1}) should be replaced by the limits of Eq.~(\ref{eq:energycon3}), whist those of the second integral should be replaced by Eq.~(\ref{eq:energycon4}). Once again neglecting derivatives of $a$, we obtain
\beq
\frac{\rm d}{{\rm d}\tau}\left(a^4\rho_\chi \right) \approx a^2\,\frac{\tri^2\Phi_0^2\mphi}{64\pi}\sqrt{1-4\delta^2_M} +a^{-1}\, \frac{(\qtc-\mphi^2/f^2)^2\Phi_0^4\mphi}{128\pi}\sqrt{1-\delta^2_M}\,,
\eeq
where we have discarded the interference term since it vanishes when averaged over time. Replacing factors of $a$ using $\rho_\phi\approx \mphi^2\Phi_0^2/(2a^3)$, we are left with the familiar Boltzmann equation:
\beq
a^{-4}\frac{\rm d}{{\rm d}t} \left(a^4\rho_\chi \right) \approx \Gamma_{\phi\to\chi\chi} \rho_\phi + 2\frac{[\sigma_{\phi\phi\to\chi\chi}v]_{v=0}}{\mphi}\rho_\phi^2\,,
\eeq
where
\beq
\Gamma_{\phi\to\chi\chi} =  \frac{\tri^2}{32\pi\mphi}\sqrt{1-4\frac{\mchi^2}{\mphi^2}}\,, \qquad
[\sigma_{\phi\phi\to\chi\chi} v]_{v=0} = \frac{(\qtc-\mphi^2/f^2)^2}{64\pi\mphi^2}\sqrt{1-\frac{\mchi^2}{\mphi^2}}\,.
\label{eq:rates}
\eeq
The decay rate $\Gamma_{\phi\to\chi\chi}$ agrees with the tree--level result obtained from QFT. The cross section $\sigma_{\phi\phi\to\chi\chi}$ also agrees with QFT so long as the Feynman amplitude is evaluated at zero relative velocity, $v=0$.\\

Note that $\phi$, as a CP odd particle, could have couplings to vector bosons as an axion. For example, it could have couplings to gluons and photons as
\bea
{\cal L}_{CP} = \frac{c_\gamma \alpha}{f} \phi \, F_{\mu\nu} \tilde{F}^{\mu\nu} +\frac{c_\g \alpha_s}{f} \phi \, Tr G_{\mu\nu} \tilde{G}^{\mu\nu} 
\eea
as well as to $W$ and $Z$ bosons. These couplings could be generated by triangle diagrams involving fermionic degrees of freedom coupled to SM gauge interactions. Whether these are present or not is a highly-model dependent question, whereas we have focused in this paper on interactions between the Goldstone bosons (the Higgs and the inflaton). We refer the reader to Refs.~\cite{Linde:2012bt,Adshead:2015pva}  for a thorough analysis of preheating due to non-zero couplings to gauge bosons.  

Conservation of energy demands $a^{-3}\frac{\rm d}{{\rm d}t} \left(a^3\rho_\phi \right)=-a^{-4}\frac{\rm d}{{\rm d}t} \left(a^4\rho_\chi \right)$, which gives
\beq
\frac{\rm d}{{\rm d}t} \left(a^3\rho_\phi \right) = - \Gamma_{\phi\to\chi\chi} \left( a^3\rho_\phi \right) - 2\frac{[\sigma_{\phi\phi\to\chi\chi}v]_{v=0}}{\mphi a^3}\left(a^3\rho_\phi\right)^2\,.
\label{eq:Boltzphi}
\eeq
If the trilinear interaction is absent ($\tri=0$) we can integrate Eq.~(\ref{eq:Boltzphi}) to show that $a^3\rho_\phi\to{\rm const}$ as $t\to\infty$. This means that the inflaton does not completely decay: volume dilution due to the Hubble expansion takes place faster than the annihilation process $\phi\phi\to\chi\chi$ can drain energy from the inflaton condensate. In order to successfully reheat the universe, the trilinear coupling must be present. Indeed, in the absence of $\phi\phi\to\chi\chi$ annihilations, (if $\qtc=\mphi^2/f^2$) we can integrate  Eq.~(\ref{eq:Boltzphi}) to show that $a^3\rho_\phi\sim e^{-\Gamma t}$: in a time of order $\Gamma_{\phi\to\chi\chi}^{-1}$ the inflaton has decayed completely. For the remainder of this section we set $\qtc=\mphi^2/f^2$ in order to place order--of--magnitude bounds on the model parameters.\\

%$v_\chi\sim 246\Gev$ is the vev of the canonically normalised higgs field. With $m_b = 4.6\Gev$, $\mchi = 125.5\Gev$ this gives $\Gamma_{\chi\to b\bar b} \sim 5\Mev$.\\

Up to this point we have neglected the decay of the higgs to the SM. The dominant channel is $\chi\to b\bar b$, with width
\beq
\Gamma_{\chi\to b\bar b} = \frac{3\mchi}{8\pi}\left(\frac{m_b}{v_\chi} \right)^2\left( 1 - \frac{4m_b^2}{\mchi^2}  \right)^{3/2}\sim 5\Mev\,.
\eeq
Since $\mchi\gg m_b$, the $b\bar b$ decay products are produced relativistically:
\beq
a^{-4}\frac{\rm d}{{\rm d}t} \left(a^4\rho_{\rm b} \right) = \Gamma_{\chi\to b\bar b}\, \rho_{\rm b}\,.
\label{eq:boltzbb}
\eeq
With $\phi\phi\to\chi\chi$ processes absent, energy conservation demands:
\beq
a^{-4}\frac{\rm d}{{\rm d}t} \left(a^4\rho_\chi \right) \approx \Gamma_{\phi\to\chi\chi} \rho_\phi - \Gamma_{\chi\to b\bar b}\, \rho_{\rm b}\,,  \qquad
a^{-3}\frac{\rm d}{{\rm d}t} \left(a^3\rho_\phi \right) = - \Gamma_{\phi\to\chi\chi} \rho_\phi\,.
\label{eq:boltzphichi}
\eeq

Eqs.~(\ref{eq:boltzbb}) and ~(\ref{eq:boltzphichi}) are the final Boltzmann equations describing perturbative reheating in the composite higgs model. The approximations involved in their derivation will begin to break down when the energy density of the decay products becomes comparable to the energy density of the inflaton condensate. Furthermore, as pointed out in~\cite{Kolb:2003ke}, and discussed in detail in~\cite{Drewes:2014pfa,Drewes:2013iaa}, $\Gamma_{\phi\to\chi\chi}$ develops a temperature dependence due to interactions (which we have not accounted for) between the decay products and the condensate. Indeed, as the decay products thermalise via scatterings and further decays,  they acquire a temperature dependent `plasma' mass $m_p(T)$ of the order $\sim\lambda T^2$, where $\lambda$ is a typical coupling constant for a particle in the plasma. The presence of these `thermal' masses prevent decay of the condensate if $\mphi^2\approx \lambda T^2$: the decay process becomes kinematically forbidden. An important consequence of these finite temperature corrections is that the reheating temperature, $T_R$ (the temperature at the onset of the radiation dominated phase) is generally higher compared to the naive estimate obtained via setting $\Gamma=H$ (see the following section).

\begin{figure}[h!]
\begin{center}
\includegraphics[width=12cm]{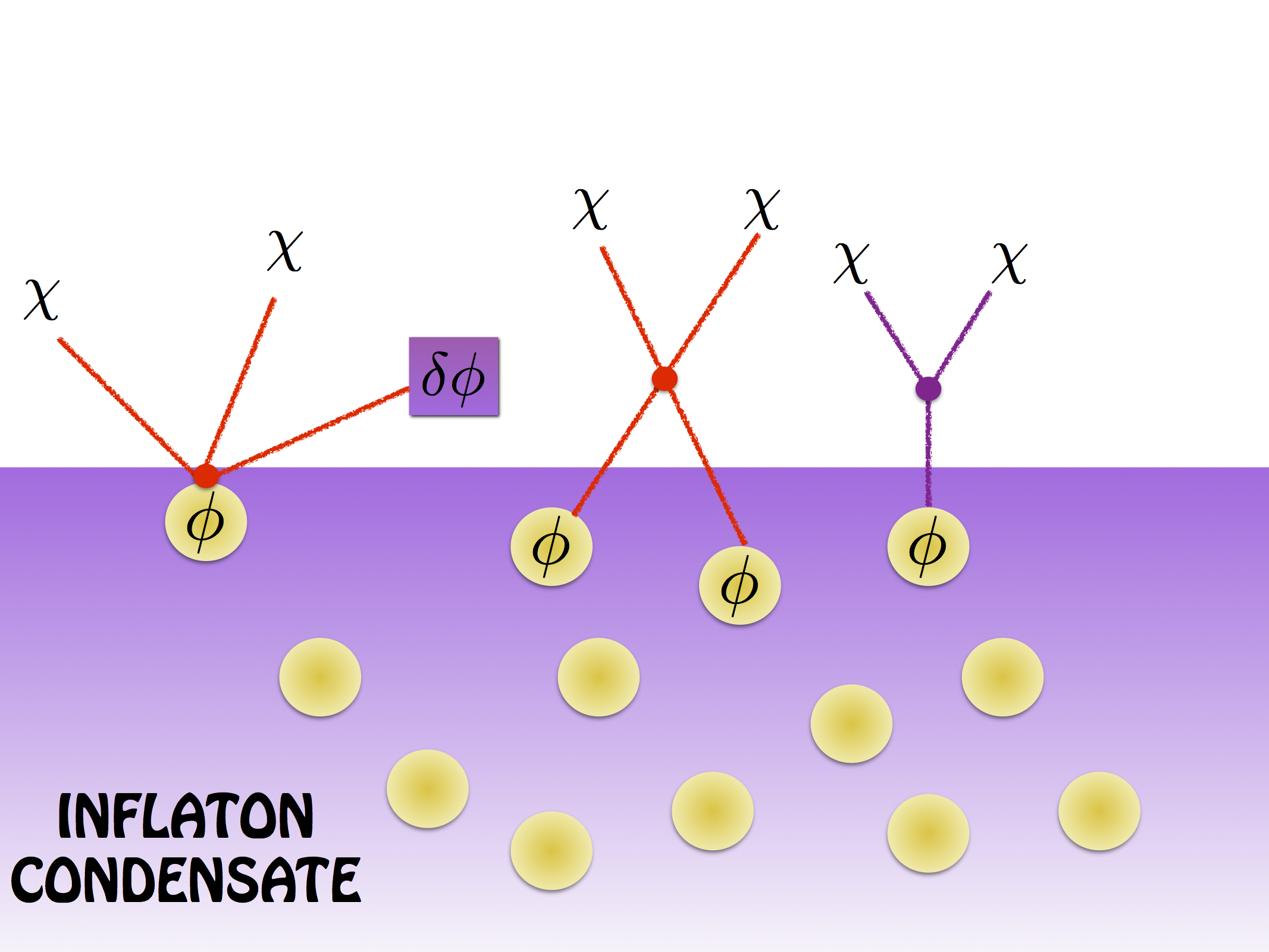}
\caption{Schematic illustration of possible inflaton--higgs interactions. The vacuum energy of the inflaton field exists as spatially coherent oscillations, which can be interpreted as a condensate of non-–relativistic zero-–momentum $\phi$--particles. The condensate decays via three--leg, $-\frac12\tri\phi\chi^2$, and four--leg, $-\frac12\qtc\phi^2\chi^2$, interactions. The Bogoliubov calculation presented in section~ \ref{sec:Bogoliubov} treats the condensate as a classical source, and so `rescattering' processes between the produced higgs particles and the condensate which excite $\delta\phi$ particles are ignored.}
\label{fig:condensate}
\end{center}
\end{figure}

In addition to the effect of thermal masses, the produced $\chi$ particles can `rescatter' off the oscillating condensate $\langle\phi\rangle$ to excite $\delta\phi$ particles. This opens another possible channel for decay of the condensate.
We illustrate this schematically in Fig.~\ref{fig:condensate} for the case of the 4--leg interaction.
In the language of our Bogoliubov calculation, this process corresponds to the term $\chi^2\phi\delta\phi$ which results from expanding $\phi$ about the mean field: $\phi(x)=\phi(t)+\delta\phi(x)$. There is also a sub--dominant process of the type $\chi\chi\to\delta\phi\delta\phi$, which is phase space suppressed. Such processes, which we have neglected in this work, will promote the decay rate $\Gamma_{\phi\to\chi\chi}$ from a constant to a function of time and temperature. To include these processes would require recourse to non--equilibrium thermal field theory, which is beyond the scope of this paper. Having acknowledged these caveats, we use the Boltzmann Equations~(\ref{eq:boltzbb}) and ~(\ref{eq:boltzphichi}) to place rough bounds on our model parameters only.

%----------------------------------------------------------------------
\subsection{Parameter Constraints from Reheating}

Combining the Planck constraint on the inflaton mass, Eq.~(\ref{eq:mphidef}), with the bound~(\ref{eq:pert}), we find that for reheating to proceed perturbatively:
\beq
\left(\frac{\tri}{\Mp}\right)^2 \ll 10^{-27}\left( \frac{f}{\Mp} \right)^2\,,
\label{eq:triBoundreheat}
\eeq
where we have used $\Phi_0\sim 0.6f$. This provides an upper bound on the trilinear coupling $\tri$ in terms of the scale $f$. A lower bound on $\tri$ can be obtained from the condition that the universe be totally radiation dominated before the BBN epoch. This requires knowledge of the reheating temperature $T_R$, which may be estimated as follows: Reheating completes at time $t_c$, when the Hubble rate $H^2=\rho/3\Mp^2\sim t_c^{-2}$ drops below the decay rate $\Gamma_{\phi\to\chi\chi} $. The density of the universe at this moment is then
\beq
\rho(t_c)\simeq3\Mp^2H^2(t_c)=3\Mp^2\Gamma_{\phi\to\chi\chi}^2\,.
\label{eq:instantdecay}
\eeq
Provided that the higgs particles are produced in thermal and chemical equilibrium, the temperature of the higgs plasma is $T_R$. Treating this ultrarelativistic gas of particles with Bose--Einstein statistics, the energy density of the universe in thermal equilibrium is then
\beq
\label{eq:BE}
\rho(T_R)\simeq\left(\frac{\pi^2}{30}\right)g_* T_R^4\,,
\eeq
where the factor $g_*(T_R)\sim10^2 -10^3$ depends on the number of ultrarelativistic degrees of freedom. Comparing Eqs.~(\ref{eq:instantdecay}) and (\ref{eq:BE}) we arrive at
\beq
T_R \approx 0.1 \sqrt{\Gamma_{\phi\to\chi\chi} \Mp}
\eeq
In order not to spoil the success of BBN, the universe must be completely dominated by relativistic particles before the BBN epoch. This constrains the reheating temperature to be $T_R\gtrsim5\Mev$~\cite{Kawasaki:1999na,Ichikawa:2005vw}, which in turn implies\footnote{We note that since $T_R$ also enters expressions for the primordial observables, the lower bound on $\Gamma_{\phi\to\chi\chi}$ given by Eq.~(\ref{eq:GammaBound}) may be tightened if our model were to be confronted with CMB data -- see for example Ref.~\cite{Martin:2014nya}.}:
\beq
\Gamma_{\phi\to\chi\chi} \gtrsim 10^{-40}\Mp\,.
\label{eq:GammaBound}
\eeq
Combining Eqs.~(\ref{eq:mphidef},\ref{eq:rates},\ref{eq:GammaBound}) we find:
\beq
\left(\frac{\tri}{\Mp}\right)^2 \gtrsim 10^{-45} \left( \frac{f}{\Mp} \right) \,.
\label{eq:TRbound}
\eeq
Finally, combining this temperature bound with the bound for perturbative reheating Eq.~(\ref{eq:triBoundreheat}), we find:
\beq
f\gg 10^{-18}\Mp\,.
\label{eq:fbound}
\eeq
%

%%%%%%%%%%%%%%%%%%%%%%%%%%%%%%%%%%%%%%%%%%%%
\section{TeV Inflaton and its consequences}
%%%%%%%%%%%%%%%%%%%%%%%%%%%%%%%%%%%%%%%%%%%% 
With the inflaton and Higgs doublet originated by the breaking of the same global symmetry, the Coleman-Weinberg contributions to their potential are naturally of the same order.  Therefore, we would expect the mass of both particles to be not far from each other, $m_\eta \sim m_h$, as well as similar size couplings. From perturbative reheating we require $m_\eta > 2 m_h$ as well as a condition on the cubic coupling Eq.(~\ref{orig:pert}), namely
\bea
\frac{c_3}{f} \ll \left(\frac{m_\eta}{f}\right)^2 \ ,
\eea
which is technically natural as the parameter $c_3$ breaks the symmetry $\eta \rightarrow - \eta$. 

Inflation would also impose a bound on the mass of the inflaton respect to the scale of breaking, see Eqs.(\ref{eq:infscale}) and (\ref{relmass}), $m_\eta/f \simeq 10^{-6} $, 
a hierarchy which is again technically natural. On the other hand, in our inflationary potential we could have added a constant term, a {\it phenomenological} cosmological constant which could change this condition and allow closer values of $f$ and $m_\eta$.

One should also keep in mind that inflation cannot last to reach energies around the MeV when the very predictive theory of Big-Bang Nucleosynthesis takes on~\cite{Hannestad:2004px}. Another constraint to keep in mind is the generation of baryon asymmetry in the Universe, which in the context of Electroweak Baryogenesis (see Ref.~\cite{Cohen:1993nk} and references therein) would require inflation to end some time before the electroweak scale. One additional attractive feature of this model is that the conditions for reheating, which in turn require CP violation, could be helpful for baryogenesis, e.g. see Ref.~\cite{Espinosa:2011eu} for a study of electroweak baryogenesis in a similar model.

If the inflaton is heavier than the Higgs doublet, one can integrate it out leading to an Effective Field Theory (EFT). In Ref.~\cite{Gorbahn:2015gxa} one can find a more general discussion on the EFT due the presence of a singlet like $\eta$, and its phenomenology. 

Interestingly, the cubic term $c_3$ is the main player in the reheating discussion as well as the collider phenomenology. The cubic term, when the Higgs acquires a vacuum expectation value $v$, would lead to a mixing of the singlet with the Higgs, resulting in two mass eigenstates with an admixture of $\eta$ and $h$. The mixing angle is given by
\bea
s_\theta \simeq \frac{c_3 v}{m_\eta^2} 
\eea

 The mixing, then, changes the way the physical SM-like Higgs behaves, as well as induces new couplings of the heavy $\eta$-like state to vector bosons and fermions.  Detailed studies from Electroweak Precision Tests (EWPT) at LEP, as well as current constraints from the measurement of the Higgs properties imposes strong bounds on this mixing. Moreover, the heavier state can be searched for directly and the reach for these searches is related to the amount of mixing. 

\begin{figure}[ht!]
\begin{center}
\includegraphics[width=0.75\textwidth]{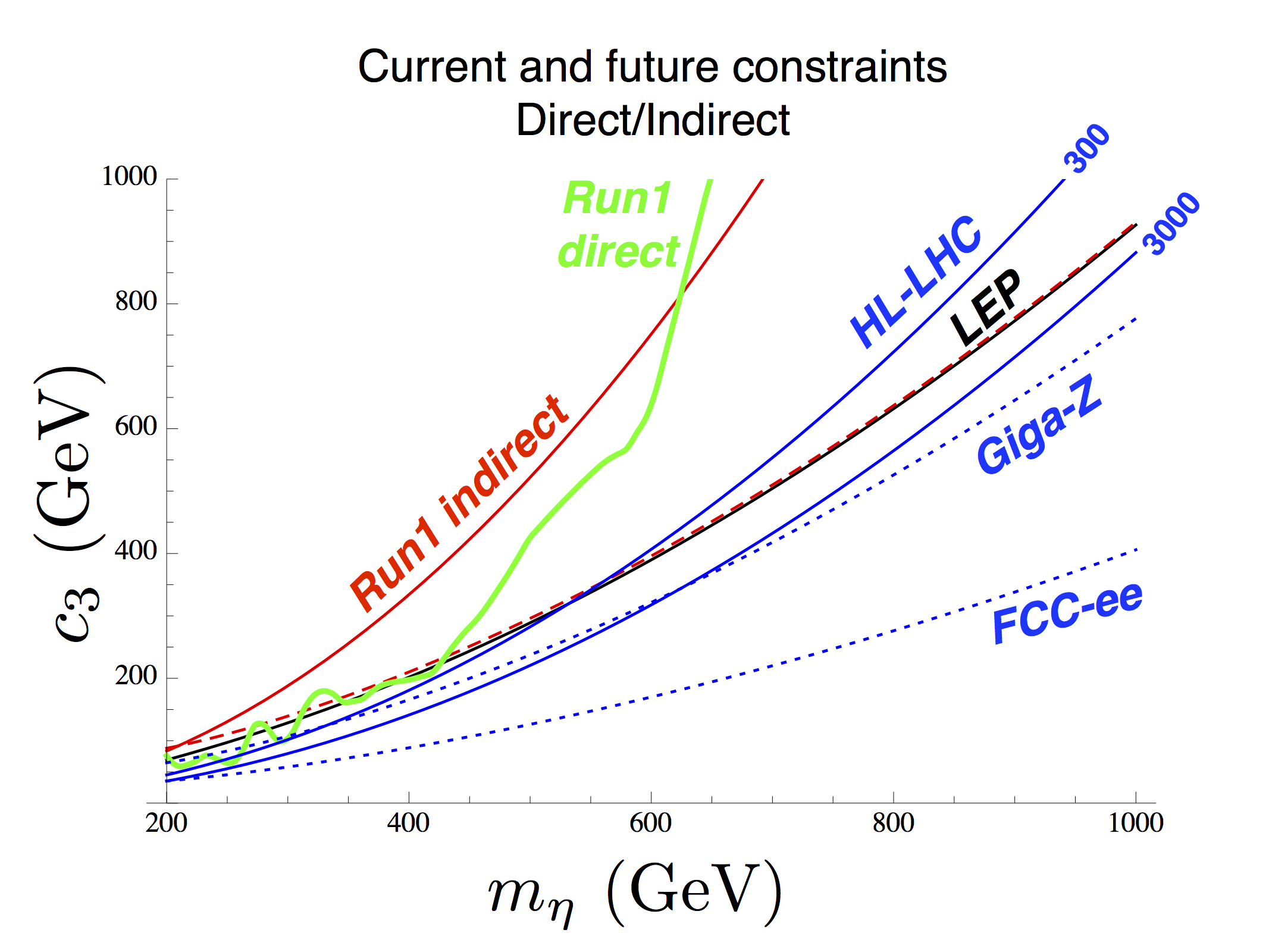} 
\caption{Present and future $95\%\, \mathrm{C.L.}$ exclusion limits in the $(m_\eta,\,c_3)$ plane  from ATLAS and CMS measurements of Higgs signal strengths (denoted by Run1 indirect) and from EWPT (denoted by LEP). Values above the red-dashed line are excluded at $95\%\, \mathrm{C.L.}$ by the combination (EWPT and Higgs signal strength). Above the green line  may also be excluded by constraints from heavy scalar searches at LHC, although these limits could be evaded in the presence of new decay modes for $\eta$. Also shown is the projected exclusion reach
from Higgs signal strengths at the 14 TeV run of LHC with $\mathcal{L}=300 \, \,\mathrm{fb}^{-1}$and at HL-LHC with $\mathcal{L}=3000 \, \,\mathrm{fb}^{-1}$ in blue, as well as projections from measurements of the $S$ and $T$ oblique parameters with ILC-{\it GigaZ} and FCC-ee in dashed-blue.}
\label{figsinglet}
\end{center}
\end{figure}

In Figure ~\ref{figsinglet}, we show current and future constraints on these parameters. They include {\it 1.)} a $\chi^2$ fit to Higgs coupling measurements~\cite{Aad:2014eha,Khachatryan:2014ira,Aad:2014eva,Chatrchyan:2013mxa,ATLAS:2013wla,Chatrchyan:2013iaa,TheATLAScollaboration:2013lia,Chatrchyan:2013zna,ATLAS_tau,Chatrchyan:2014nva}, {\it 2.)} The $95\%\, \mathrm{C.L.}$ exclusion prospects for LHC at 14 TeV with $\mathcal{L}=300 \, \,\mathrm{fb}^{-1}$  and $\mathcal{L}=3000\, \, \mathrm{fb}^{-1}$, by assuming that future measurements of Higgs signal strengths will be centered at the SM value, and use the projected CMS sensitivities, {\it 3.)} A fit to the oblique parameters $S, T, U$ using the best-fit values and standard deviations from the global analysis of the 
GFitter Group \cite{Baak:2014ora},  and finally {\it 4.)} Future limits  on EW precision observables from $e^+e^-$ colliders (see {\it e.g.} \cite{Fan:2014vta}), ILC and FCC-ee. 

The corrections to $S$ and $T$ from the inflaton-Higgs  mixing given by
\bea
\label{ST}
\Delta S = \frac{1}{\pi}\,s_\theta^2 \left[ - H_S\left(\frac{m^2_h}{m^2_Z}\right) + H_S\left(\frac{m^2_\eta}{m^2_Z}\right)\right] \nonumber \\
\Delta T = \frac{g^2}{16 \,\pi^2 \,\cW^2 \,\alpha_{\mathrm{EM}}}\,s_\theta^2 \left[ - H_T\left(\frac{m^2_h}{m^2_Z}\right) + H_T\left(\frac{m^2_\eta}{m^2_Z}\right)\right]
\eea
with the functions $H_S(x)$ and $H_T(x)$ defined in Appendix C of \cite{Hagiwara:1994pw}.

Regarding future colliders, we assumed a SM best-fit value, and interpreted the ILC {\it GigaZ} program's expected precision is $\sigma_S = 0.017$ and $\sigma_T = 0.022$ \cite{Baak:2014ora,Asner:2013psa} and the FCC-ee prospects of 
$\sigma_S = 0.007$ and $\sigma_T = 0.004$ \cite{Gomez-Ceballos:2013zzn}.  As one can see, colliders are sensitive to relatively large values of the triple coupling, whereas perturbative reheating is sensitive to lower values of the coupling.
 
Finally, note that in the explicit CP breaking scenario, there would be direct couplings of the inflaton to SM fermions ($\epsilon_f$) an these would be proportional to $c_3$, see Eq.~\ref{c3epsilon}.

%%%%%%%%%%%%%%%%%%%%%%%%%%%%%%%%%%%%%%%%%%%%
\section{Conclusions}
We have presented a single model that can realise inflation, perturbative reheating, and electroweak symmetry breaking in a natural way. In the minimal model the five Goldstone bosons from the global symmetry breaking $SO(6)\sim SU(4) \rightarrow SO(5) \sim Sp(4)$ play the role of a Higgs doublet and an inflaton singlet. We have argued that a trilinear coupling between the latter ($\eta$) and two Higgs bosons ($h$) is necessary for successful reheating, and shown under which condition this term can be present. In particular, the model needs to have broken CP, which can be realised spontaneously or explicitly. A detailed derivation of the scalar potential for $h$ and $\eta$ arising from loops of $SU(2)$ gauge bosons and fermions in the \textbf{6} of $SU(4)$ was given in the first section.

The CMB results \cite{Ade:2015lrj} allow us relate the parameters in our model, and explain mass hierarchies. A range of energy scales for inflation, or equivalently for the mass of the inflaton was presented in the second section. To the merit of the model, none of the relevant scales are expected to be affected by quantum gravity. 

The motive of perturbative reheating further fixes the parameters in the potential. For a particular range of parameter space (given by Eq. \eqref{orig:pert}) parametric instability is not triggered and non-perturbative effects are subdominant. With a Bogoliubov calculation \cite{Braden:2010wd} we find the single particle occupation numbers, and as usual the evolution of the fields is established using Boltzmann equations. We finished this section by an exposition of the numerical constraints on the reheating temperature and the model parameters from perturbativity (\ref{eq:GammaBound}-\ref{eq:fbound}).

We have also explored the possibility of TeV values of the inflaton mass and coupling to the Higgs. As an effective theory, the inflaton's effect at low energies is inducing a mixing effect in the Higgs particle properties, an effect which is constrained by precise electroweak data as well as the LHC. We discussed the future reach for colliders on the inflaton-Higgs parameter space, finding that while perturbative reheating explores a region of small mixing, colliders are most sensitive to large values of this parameter. 

The model building presented in this paper hints at interesting opportunities for further studies. The fact that the model is able to address and connect normally unrelated cosmological events in a natural way makes that the considerations here may indeed tempt the reader to further inquiry, in the light of recent developments. As mentioned in the introduction, the discussion of cosmological relaxation by an interplay between the Higgs and a pGB \cite{Graham:2015cka} offers an attractive example. Other directions include an investigation of the changed evolution of the Higgs dynamics and its implications on electroweak stability \cite{Gross:2015bea}, possible UV completions for which the present theory is a boundary condition at low energy (on which we commented in \cite{Croon:2015fza}), as well as the implications of CP violation and the inflaton degree of freedom for electroweak baryonenesis.

\section*{Acknowledgements}
We would like to thank Marco Drewes, Hitoshi Murayama and Josemi No for discussions.  
This work  is supported by the Science Technology and Facilities Council (STFC) under grant number ST/L000504/1. 

\begin{appendices}
\section{Computation of the scalar potential}
\subsection{Composite Higgs vacuum}
At one loop, the Coleman-Weinberg potential due to up-type quarks coupling to $\Sigma$ as in \eqref{fermionL} is given by\footnote{In general there will be contributions from down type quarks and gauge bosons as well. In fact, it should be noted that at least one other fermion generation is needed to make the CP assignment physical \cite{Redi:2012ha}. However, these will not lead to different couplings in the scalar potential, and here we take them to be sub-leading corrections to the coefficients.}
\bea V(h,\eta) = - 2 N_c \int \frac{d^4 p}{(2 \pi)^4} \log \left( p^2 \Pi_{u_L}\Pi_{u_R} - \left|\Pi_{u_L u_R} \right|^2 \right) \eea
where we have used new form factors for simplicity, which are just rotations of the original parameters in the Lagrangian \eqref{fermionL}:
\begin{subequations}
\begin{align} \Pi_{u_L} &= \frac{\Pi^q_0 + \Pi_0^{q'}}{2} - \Pi_1^q\, \frac{\text{Tr}[\bar\Psi_q \Sigma ]\slashed{p}\text{Tr}[\Psi_q \Sigma^{\dagger}]}{\bar{u}_L \slashed{p} u_L}\\ 
\Pi_{u_R} &= \Pi_0^u- \Pi_1^u \, \frac{\text{Tr}[\bar\Psi_u \Sigma ] \slashed{p} \text{Tr}[\Psi_u \Sigma^{\dagger}]}{\bar{u}_R \slashed{p} u_R} \\ 
\Pi_{u_L u_R} &= M_1^u \, \frac{\text{Tr}[\bar\Psi_q \Sigma ] \text{Tr}[\Psi_u \Sigma^{\dagger}]}{\bar{u}_L  u_R}
\end{align}
\end{subequations}
as explained in the main text, we refer to $\Psi$ as the fermion multiplets in the 6 of SU(4). 

If we assume the ratios form factors fall off rapidly enough with momentum to make the integrals converge, we may expand the logarithms to find the following Lagrangian to fourth order in the fields:\footnote{This is a common assumption, motivated by the fact that higher order terms are expected to be suppressed by squares of ratios of form factors. In other words, this falls under the same assumption as the convergence of the integrals.} 
\bea V(\phi,h) = a_1 h^2 + \lambda h^4 + |\kappa|^2 \left( a_2 + a_3 h^2 + a_4 |\kappa|^2 \right) \label{CHpot}
\eea
where $\kappa = \sqrt{f^2-h^2 -\eta^2} + i \epsilon_t \eta$. The coefficients are given by integrals over the form factors of the fields contributing to the CW potential: the gauge bosons, and the up-type and down-type fermions.
If we assume the contributions are dominated by the heaviest up-type quark, which we will call the top as in the Standard Model (while this quark is not necessarily identified with the Standard Model top), the coefficients are given by:
\begin{subequations}
\begin{align}
a_1 &=- 2 f^2 N_c \int \frac{d^4 p}{(2 \pi)^4} \frac{1}{\Pi_0}\,\, \left(  -4 \Pi^q_1 \Pi^t_0 \right) \\
a_2 &=- 2 f^2 N_c \int \frac{d^4 p}{(2 \pi)^4} \frac{1}{\Pi_0}\,\, \left(  -2 \Pi^q_0 \Pi_1^{t}-2 \Pi^{q'}_1 \Pi_1^{t} \right) \\
a_3 &=- 2 N_c \int \frac{d^4 p}{(2 \pi)^4} \frac{1}{\Pi_0}\,\, \left( -\frac{4 |M_t^1|^2 }{p^2}+\frac{8 \Pi^q_0 \Pi^q_1 \Pi^t_0 \Pi^t_1 }{\Pi_0}+\frac{8 \Pi^q_1 \Pi^{q'}_1 \Pi^t_0 \Pi^t_1  }{\Pi_0}+16 \Pi^q_1 \Pi^t_1  \right) \\
a_4 &= - 2 N_c \int \frac{d^4 p}{(2 \pi)^4} \frac{1}{\Pi_0}\,\, \left(  \frac{2 (\Pi^q_0)^2 (\Pi^t_1)^2}{\Pi_0}+\frac{4 \Pi^q_0 \Pi^{q'}_1 (\Pi^t_1)^2}{\Pi_0}+\frac{2 (\Pi^{q'}_1)^2 (\Pi^t_1)^2}{\Pi_0} \right) \\
\lambda &= - 2 N_c \int \frac{d^4 p}{(2 \pi)^4} \frac{1}{\Pi_0}\,\, \left( \frac{8( \Pi^q_1)^2 (\Pi^t_0)^2}{\Pi_0} \right)
\end{align}
\end{subequations}
where $\Pi_0 $ is the relevant field independent factor:
\bea \Pi_0 = \frac{1}{2} \Pi_0^t \left( \Pi_0^q + \Pi_0^{q'}\right)  \eea 
i.e., a function of the different propagation terms for the fermions, the first terms in the fermion Lagrangian \eqref{fermionL}. Also, note we have defined
\bea
p \to p/f 
\eea
for simplicity.

\subsection{CP breaking vacuum}

Here we repeat the exercise in the previous section to compute the coefficients of the CP breaking vacuum potential,
\bea V(\eta, h) =  c_{tad} \eta + m_\eta^2 \eta^2  +\tilde{c}_\eta \eta^3 + \lambda_\eta \eta^4 +m_h^2 h^2  + \lambda_h h^4  + c_3 \eta  h^2  + c_4 \eta^2 h^2 
\eea
The coefficients $c_i$ are in general nonzero, except for at $\alpha = 1/4 \pi$. Below we compute the parameters in an example with $\alpha = 1/3 \pi$ case. As argued in the main text, the $\alpha = 2/3 \pi$ case can be obtained from this by making the substitution $\eta \rightarrow - \eta$ in the potential:\footnote{These are again the parameters before shifting away the tadpole term, in exactly the same way as above.}
\begin{subequations}
\begin{align}
c_{tad} &=- 2 N_c f^3 \int \frac{d^4 p}{(2 \pi)^4}  \frac{1}{2 \Pi_0} \sqrt{3} \eta  \Pi_1^t ((\epsilon_u -4)\epsilon_u  -1)    (\Pi_0^q+\Pi_0^{q'})\\
m_\eta^2 &=- 2 N_c f^2 \int \frac{d^4 p}{(2 \pi)^4} \frac{1}{\Pi_0} \left(  \frac{3 \Pi_1^t (( \epsilon_u-4)\epsilon_u  -1)^2  (\Pi_0^q+\Pi_0^{q'})^2}{8 \Pi_0}- \Pi_1^t \left( \epsilon_u^2-1\right) (\Pi_0^q+\Pi_0^{q'}) ) \right) \\
\tilde{c}_\eta &=- 2 N_c f \int \frac{d^4 p}{(2 \pi)^4} \left( -\frac{\sqrt{3} \eta ^3 (\Pi_1^t)^2  ( \epsilon_u^2-1) ((\epsilon_u -4)\epsilon_u  -1)    (\Pi_0^q+\Pi_0^{q'}) \left( \Pi_0^q+\Pi_0^{q'}\right)}{2 \Pi_0^2} \right)  \\
\lambda_\eta  &= - 2 N_c \int \frac{d^4 p}{(2 \pi)^4} \frac{ (\Pi_1^t)^2 \left( \epsilon_u^2-1\right)^2 \left(\Pi_0^q+\Pi_0^{q'}\right)^2}{2 \Pi_0^2} \\
m_h^2 &=  - 2 N_c \int \frac{d^4 p}{(2 \pi)^4} \frac{ \left(p^2 \left(\Pi_1^t \left( \epsilon_u^2+3\right) (\Pi_0^q+2 \Pi_1^q+\Pi_0^{q'})-2 \Pi_1^q \Pi_0^t\right)-2 M_t^2 \left( \epsilon_u^2+3\right)\right)}{2 p^2 \Pi_0}   \\
\lambda_h &= - 2 N_c f^2 \int \frac{d^4 p}{(2 \pi)^4} \frac{ \left(\epsilon_u^2+3\right) \left(M_q^2 -p^2 \Pi_1^q \Pi_1^t \right)}{p^2 \Pi_0} +\frac{\left(\Pi_1^t \left(\epsilon_u^2+3\right) (\Pi_0^q+2 \Pi_1^q+\Pi_0^{q'})-2 \Pi_1^q \Pi_0^t \right)^2}{8\Pi_0^2} \\ 
c_3 &= - 2 N_c f \int \frac{d^4 p}{(2 \pi)^4}  \left( \frac{\sqrt{3}  (( \epsilon_u-4) \epsilon_u -1)    \left(M_t^2-p^2 \Pi_1^q \Pi_1^t\right)}{p^2 \Pi_0} \right. \\ & \left. +\frac{\sqrt{3} \Pi_1^t (( \epsilon_u-4) \epsilon_u -1)    (\Pi_0^q+\Pi_0^{q'}) \left(\Pi_1^t \left( \epsilon_u^2+3\right) (\Pi_0^q+2 \Pi_1^q+\Pi_0^{q'})-2 \Pi_1^q \Pi_0^t\right)}{4 \Pi_0^2}  \right) \end{align}
\begin{align}
c_4 &= - 2 N_c \int \frac{d^4 p}{(2 \pi)^4} \left( \frac{2  \left( \epsilon_u^2-1\right) \left(p^2 \Pi_1^q \Pi_1^t-M_t^2\right)}{p^2 \Pi_0} \right. \\ & \left. -\frac{\eta ^2 h^2 \Pi_1^t \left( \epsilon_u^2-1\right) (\Pi_0^q+\Pi_0^{q'}) \left(\Pi_1^t \left( \epsilon_u^2+3\right) (\Pi_0^q+2 \Pi_1^q+\Pi_0^{q'})-2 \Pi_1^q \Pi_0^t\right)}{2 \Pi_0^2} \right)
\end{align}
\end{subequations}
where again $\Pi_0 $ is the relevant field independent factor, here given by:
\bea \Pi_0 = \frac{1}{2} (\Pi_0^{q}+\Pi_0^{q'}) \left(\Pi_0^t-2  \Pi_1^t  \left( \epsilon_u^2+1\right)\right)\eea
As explained in the main text, the tadpole term can be shifted away by an appropriate shift in the other parameters, corresponding to a vev for $\eta$: 
\bea \notag c_{tad} + 2 m_\eta^2 v_\eta + 3 \tilde{c}_\eta v_\eta^2 + 4 \lambda_\eta v_\eta^3 = 0 \eea
The new parameters will then be given in terms of the quoted parameters as 
\begin{subequations}
\begin{align} m_\eta^2 &\rightarrow  m_\eta^2 + 3 \tilde{c}_\eta v_\eta + 6 \lambda_\eta v_\eta^2 \\ 
\tilde{c}_\eta &\rightarrow \tilde{c}_\eta + 4 v_\eta \lambda_\eta \\ 
m_h^2 &\rightarrow c_3 v_\eta + c_4 v_\eta^2 \\
c_3 &\rightarrow c_3 + 2 c_4 v_\eta  \end{align}
\end{subequations}

\end{appendices}

%\section*{Bibliography}
\bibliographystyle{apsrev4-1}
\bibliography{bib_file}

\begin{thebibliography}{55}%
\makeatletter
\providecommand \@ifxundefined [1]{%
 \@ifx{#1\undefined}
}%
\providecommand \@ifnum [1]{%
 \ifnum #1\expandafter \@firstoftwo
 \else \expandafter \@secondoftwo
 \fi
}%
\providecommand \@ifx [1]{%
 \ifx #1\expandafter \@firstoftwo
 \else \expandafter \@secondoftwo
 \fi
}%
\providecommand \natexlab [1]{#1}%
\providecommand \enquote  [1]{``#1''}%
\providecommand \bibnamefont  [1]{#1}%
\providecommand \bibfnamefont [1]{#1}%
\providecommand \citenamefont [1]{#1}%
\providecommand \href@noop [0]{\@secondoftwo}%
\providecommand \href [0]{\begingroup \@sanitize@url \@href}%
\providecommand \@href[1]{\@@startlink{#1}\@@href}%
\providecommand \@@href[1]{\endgroup#1\@@endlink}%
\providecommand \@sanitize@url [0]{\catcode `\\12\catcode `\$12\catcode
  `\&12\catcode `\#12\catcode `\^12\catcode `\_12\catcode `\%12\relax}%
\providecommand \@@startlink[1]{}%
\providecommand \@@endlink[0]{}%
\providecommand \url  [0]{\begingroup\@sanitize@url \@url }%
\providecommand \@url [1]{\endgroup\@href {#1}{\urlprefix }}%
\providecommand \urlprefix  [0]{URL }%
\providecommand \Eprint [0]{\href }%
\providecommand \doibase [0]{http://dx.doi.org/}%
\providecommand \selectlanguage [0]{\@gobble}%
\providecommand \bibinfo  [0]{\@secondoftwo}%
\providecommand \bibfield  [0]{\@secondoftwo}%
\providecommand \translation [1]{[#1]}%
\providecommand \BibitemOpen [0]{}%
\providecommand \bibitemStop [0]{}%
\providecommand \bibitemNoStop [0]{.\EOS\space}%
\providecommand \EOS [0]{\spacefactor3000\relax}%
\providecommand \BibitemShut  [1]{\csname bibitem#1\endcsname}%
\let\auto@bib@innerbib\@empty
%</preamble>
\bibitem [{\citenamefont {Lyth}(1997)}]{Lyth:1996im}%
  \BibitemOpen
  \bibfield  {author} {\bibinfo {author} {\bibfnamefont {D.~H.}\ \bibnamefont
  {Lyth}},\ }\href {\doibase 10.1103/PhysRevLett.78.1861} {\bibfield  {journal}
  {\bibinfo  {journal} {Phys. Rev. Lett.}\ }\textbf {\bibinfo {volume} {78}},\
  \bibinfo {pages} {1861} (\bibinfo {year} {1997})},\ \Eprint
  {http://arxiv.org/abs/hep-ph/9606387} {arXiv:hep-ph/9606387 [hep-ph]}
  \BibitemShut {NoStop}%
%%CITATION = HEP-PH/9606387;%%
\bibitem [{\citenamefont {Agashe}\ \emph {et~al.}(2005)\citenamefont {Agashe},
  \citenamefont {Contino},\ and\ \citenamefont {Pomarol}}]{Agashe:2004rs}%
  \BibitemOpen
  \bibfield  {author} {\bibinfo {author} {\bibfnamefont {K.}~\bibnamefont
  {Agashe}}, \bibinfo {author} {\bibfnamefont {R.}~\bibnamefont {Contino}}, \
  and\ \bibinfo {author} {\bibfnamefont {A.}~\bibnamefont {Pomarol}},\ }\href
  {\doibase 10.1016/j.nuclphysb.2005.04.035} {\bibfield  {journal} {\bibinfo
  {journal} {Nucl. Phys.}\ }\textbf {\bibinfo {volume} {B719}},\ \bibinfo
  {pages} {165} (\bibinfo {year} {2005})},\ \Eprint
  {http://arxiv.org/abs/hep-ph/0412089} {arXiv:hep-ph/0412089 [hep-ph]}
  \BibitemShut {NoStop}%
%%CITATION = HEP-PH/0412089;%%
\bibitem [{\citenamefont {Bellazzini}\ \emph {et~al.}(2014)\citenamefont
  {Bellazzini}, \citenamefont {Cs­Ìi},\ and\ \citenamefont
  {Serra}}]{Bellazzini:2014yua}%
  \BibitemOpen
  \bibfield  {author} {\bibinfo {author} {\bibfnamefont {B.}~\bibnamefont
  {Bellazzini}}, \bibinfo {author} {\bibfnamefont {C.}~\bibnamefont {Cs­Ìi}}, \
  and\ \bibinfo {author} {\bibfnamefont {J.}~\bibnamefont {Serra}},\ }\href
  {\doibase 10.1140/epjc/s10052-014-2766-x} {\bibfield  {journal} {\bibinfo
  {journal} {Eur. Phys. J.}\ }\textbf {\bibinfo {volume} {C74}},\ \bibinfo
  {pages} {2766} (\bibinfo {year} {2014})},\ \Eprint
  {http://arxiv.org/abs/1401.2457} {arXiv:1401.2457 [hep-ph]} \BibitemShut
  {NoStop}%
%%CITATION = ARXIV:1401.2457;%%
\bibitem [{\citenamefont {Freese}\ \emph {et~al.}(1990)\citenamefont {Freese},
  \citenamefont {Frieman},\ and\ \citenamefont {Olinto}}]{Freese:1990rb}%
  \BibitemOpen
  \bibfield  {author} {\bibinfo {author} {\bibfnamefont {K.}~\bibnamefont
  {Freese}}, \bibinfo {author} {\bibfnamefont {J.~A.}\ \bibnamefont {Frieman}},
  \ and\ \bibinfo {author} {\bibfnamefont {A.~V.}\ \bibnamefont {Olinto}},\
  }\href {\doibase 10.1103/PhysRevLett.65.3233} {\bibfield  {journal} {\bibinfo
   {journal} {Phys. Rev. Lett.}\ }\textbf {\bibinfo {volume} {65}},\ \bibinfo
  {pages} {3233} (\bibinfo {year} {1990})}\BibitemShut {NoStop}%
%%CITATION = PRLTA,65,3233;%%
\bibitem [{\citenamefont {Arkani-Hamed}\ \emph
  {et~al.}(2003{\natexlab{a}})\citenamefont {Arkani-Hamed}, \citenamefont
  {Cheng}, \citenamefont {Creminelli},\ and\ \citenamefont
  {Randall}}]{ArkaniHamed:2003wu}%
  \BibitemOpen
  \bibfield  {author} {\bibinfo {author} {\bibfnamefont {N.}~\bibnamefont
  {Arkani-Hamed}}, \bibinfo {author} {\bibfnamefont {H.-C.}\ \bibnamefont
  {Cheng}}, \bibinfo {author} {\bibfnamefont {P.}~\bibnamefont {Creminelli}}, \
  and\ \bibinfo {author} {\bibfnamefont {L.}~\bibnamefont {Randall}},\ }\href
  {\doibase 10.1103/PhysRevLett.90.221302} {\bibfield  {journal} {\bibinfo
  {journal} {Phys. Rev. Lett.}\ }\textbf {\bibinfo {volume} {90}},\ \bibinfo
  {pages} {221302} (\bibinfo {year} {2003}{\natexlab{a}})},\ \Eprint
  {http://arxiv.org/abs/hep-th/0301218} {arXiv:hep-th/0301218 [hep-th]}
  \BibitemShut {NoStop}%
%%CITATION = HEP-TH/0301218;%%
\bibitem [{\citenamefont {Linde}(1994)}]{Linde:1993cn}%
  \BibitemOpen
  \bibfield  {author} {\bibinfo {author} {\bibfnamefont {A.~D.}\ \bibnamefont
  {Linde}},\ }\href {\doibase 10.1103/PhysRevD.49.748} {\bibfield  {journal}
  {\bibinfo  {journal} {Phys. Rev.}\ }\textbf {\bibinfo {volume} {D49}},\
  \bibinfo {pages} {748} (\bibinfo {year} {1994})},\ \Eprint
  {http://arxiv.org/abs/astro-ph/9307002} {arXiv:astro-ph/9307002 [astro-ph]}
  \BibitemShut {NoStop}%
%%CITATION = ASTRO-PH/9307002;%%
\bibitem [{\citenamefont {Kim}\ \emph {et~al.}(2005)\citenamefont {Kim},
  \citenamefont {Nilles},\ and\ \citenamefont {Peloso}}]{Kim:2004rp}%
  \BibitemOpen
  \bibfield  {author} {\bibinfo {author} {\bibfnamefont {J.~E.}\ \bibnamefont
  {Kim}}, \bibinfo {author} {\bibfnamefont {H.~P.}\ \bibnamefont {Nilles}}, \
  and\ \bibinfo {author} {\bibfnamefont {M.}~\bibnamefont {Peloso}},\ }\href
  {\doibase 10.1088/1475-7516/2005/01/005} {\bibfield  {journal} {\bibinfo
  {journal} {JCAP}\ }\textbf {\bibinfo {volume} {0501}},\ \bibinfo {pages}
  {005} (\bibinfo {year} {2005})},\ \Eprint
  {http://arxiv.org/abs/hep-ph/0409138} {arXiv:hep-ph/0409138 [hep-ph]}
  \BibitemShut {NoStop}%
%%CITATION = HEP-PH/0409138;%%
\bibitem [{\citenamefont {Dimopoulos}\ \emph {et~al.}(2008)\citenamefont
  {Dimopoulos}, \citenamefont {Kachru}, \citenamefont {McGreevy},\ and\
  \citenamefont {Wacker}}]{Dimopoulos:2005ac}%
  \BibitemOpen
  \bibfield  {author} {\bibinfo {author} {\bibfnamefont {S.}~\bibnamefont
  {Dimopoulos}}, \bibinfo {author} {\bibfnamefont {S.}~\bibnamefont {Kachru}},
  \bibinfo {author} {\bibfnamefont {J.}~\bibnamefont {McGreevy}}, \ and\
  \bibinfo {author} {\bibfnamefont {J.~G.}\ \bibnamefont {Wacker}},\ }\href
  {\doibase 10.1088/1475-7516/2008/08/003} {\bibfield  {journal} {\bibinfo
  {journal} {JCAP}\ }\textbf {\bibinfo {volume} {0808}},\ \bibinfo {pages}
  {003} (\bibinfo {year} {2008})},\ \Eprint
  {http://arxiv.org/abs/hep-th/0507205} {arXiv:hep-th/0507205 [hep-th]}
  \BibitemShut {NoStop}%
%%CITATION = HEP-TH/0507205;%%
\bibitem [{\citenamefont {Copeland}\ \emph {et~al.}(1999)\citenamefont
  {Copeland}, \citenamefont {Mazumdar},\ and\ \citenamefont
  {Nunes}}]{Copeland:1999cs}%
  \BibitemOpen
  \bibfield  {author} {\bibinfo {author} {\bibfnamefont {E.~J.}\ \bibnamefont
  {Copeland}}, \bibinfo {author} {\bibfnamefont {A.}~\bibnamefont {Mazumdar}},
  \ and\ \bibinfo {author} {\bibfnamefont {N.~J.}\ \bibnamefont {Nunes}},\
  }\href {\doibase 10.1103/PhysRevD.60.083506} {\bibfield  {journal} {\bibinfo
  {journal} {Phys. Rev.}\ }\textbf {\bibinfo {volume} {D60}},\ \bibinfo {pages}
  {083506} (\bibinfo {year} {1999})},\ \Eprint
  {http://arxiv.org/abs/astro-ph/9904309} {arXiv:astro-ph/9904309 [astro-ph]}
  \BibitemShut {NoStop}%
%%CITATION = ASTRO-PH/9904309;%%
\bibitem [{\citenamefont {Silverstein}\ and\ \citenamefont
  {Westphal}(2008)}]{Silverstein:2008sg}%
  \BibitemOpen
  \bibfield  {author} {\bibinfo {author} {\bibfnamefont {E.}~\bibnamefont
  {Silverstein}}\ and\ \bibinfo {author} {\bibfnamefont {A.}~\bibnamefont
  {Westphal}},\ }\href {\doibase 10.1103/PhysRevD.78.106003} {\bibfield
  {journal} {\bibinfo  {journal} {Phys. Rev.}\ }\textbf {\bibinfo {volume}
  {D78}},\ \bibinfo {pages} {106003} (\bibinfo {year} {2008})},\ \Eprint
  {http://arxiv.org/abs/0803.3085} {arXiv:0803.3085 [hep-th]} \BibitemShut
  {NoStop}%
%%CITATION = ARXIV:0803.3085;%%
\bibitem [{\citenamefont {Arkani-Hamed}\ \emph
  {et~al.}(2003{\natexlab{b}})\citenamefont {Arkani-Hamed}, \citenamefont
  {Cheng}, \citenamefont {Creminelli},\ and\ \citenamefont
  {Randall}}]{ArkaniHamed:2003mz}%
  \BibitemOpen
  \bibfield  {author} {\bibinfo {author} {\bibfnamefont {N.}~\bibnamefont
  {Arkani-Hamed}}, \bibinfo {author} {\bibfnamefont {H.-C.}\ \bibnamefont
  {Cheng}}, \bibinfo {author} {\bibfnamefont {P.}~\bibnamefont {Creminelli}}, \
  and\ \bibinfo {author} {\bibfnamefont {L.}~\bibnamefont {Randall}},\ }\href
  {\doibase 10.1088/1475-7516/2003/07/003} {\bibfield  {journal} {\bibinfo
  {journal} {JCAP}\ }\textbf {\bibinfo {volume} {0307}},\ \bibinfo {pages}
  {003} (\bibinfo {year} {2003}{\natexlab{b}})},\ \Eprint
  {http://arxiv.org/abs/hep-th/0302034} {arXiv:hep-th/0302034 [hep-th]}
  \BibitemShut {NoStop}%
%%CITATION = HEP-TH/0302034;%%
\bibitem [{\citenamefont {Croon}\ and\ \citenamefont
  {Sanz}(2015)}]{Croon:2014dma}%
  \BibitemOpen
  \bibfield  {author} {\bibinfo {author} {\bibfnamefont {D.}~\bibnamefont
  {Croon}}\ and\ \bibinfo {author} {\bibfnamefont {V.}~\bibnamefont {Sanz}},\
  }\href {\doibase 10.1088/1475-7516/2015/02/008} {\bibfield  {journal}
  {\bibinfo  {journal} {JCAP}\ }\textbf {\bibinfo {volume} {1502}},\ \bibinfo
  {pages} {008} (\bibinfo {year} {2015})},\ \Eprint
  {http://arxiv.org/abs/1411.7809} {arXiv:1411.7809 [hep-ph]} \BibitemShut
  {NoStop}%
%%CITATION = ARXIV:1411.7809;%%
\bibitem [{\citenamefont {Croon}\ \emph {et~al.}(2015)\citenamefont {Croon},
  \citenamefont {Sanz},\ and\ \citenamefont {Setford}}]{Croon:2015fza}%
  \BibitemOpen
  \bibfield  {author} {\bibinfo {author} {\bibfnamefont {D.}~\bibnamefont
  {Croon}}, \bibinfo {author} {\bibfnamefont {V.}~\bibnamefont {Sanz}}, \ and\
  \bibinfo {author} {\bibfnamefont {J.}~\bibnamefont {Setford}},\ }\href@noop
  {} {\  (\bibinfo {year} {2015})},\ \Eprint {http://arxiv.org/abs/1503.08097}
  {arXiv:1503.08097 [hep-ph]} \BibitemShut {NoStop}%
%%CITATION = ARXIV:1503.08097;%%
\bibitem [{\citenamefont {Ade}\ \emph {et~al.}(2015)\citenamefont {Ade} \emph
  {et~al.}}]{Ade:2015lrj}%
  \BibitemOpen
  \bibfield  {author} {\bibinfo {author} {\bibfnamefont {P.}~\bibnamefont
  {Ade}} \emph {et~al.} (\bibinfo {collaboration} {Planck Collaboration}),\
  }\href@noop {} {\  (\bibinfo {year} {2015})},\ \Eprint
  {http://arxiv.org/abs/1502.02114} {arXiv:1502.02114 [astro-ph.CO]}
  \BibitemShut {NoStop}%
%%CITATION = ARXIV:1502.02114;%%
\bibitem [{\citenamefont {Graham}\ \emph {et~al.}(2015)\citenamefont {Graham},
  \citenamefont {Kaplan},\ and\ \citenamefont {Rajendran}}]{Graham:2015cka}%
  \BibitemOpen
  \bibfield  {author} {\bibinfo {author} {\bibfnamefont {P.~W.}\ \bibnamefont
  {Graham}}, \bibinfo {author} {\bibfnamefont {D.~E.}\ \bibnamefont {Kaplan}},
  \ and\ \bibinfo {author} {\bibfnamefont {S.}~\bibnamefont {Rajendran}},\
  }\href@noop {} {\  (\bibinfo {year} {2015})},\ \Eprint
  {http://arxiv.org/abs/1504.07551} {arXiv:1504.07551 [hep-ph]} \BibitemShut
  {NoStop}%
%%CITATION = ARXIV:1504.07551;%%
\bibitem [{\citenamefont {Espinosa}\ \emph {et~al.}(2015)\citenamefont
  {Espinosa}, \citenamefont {Grojean}, \citenamefont {Panico}, \citenamefont
  {Pomarol}, \citenamefont {Pujol¯Ô},\ and\ \citenamefont
  {Servant}}]{Espinosa:2015eda}%
  \BibitemOpen
  \bibfield  {author} {\bibinfo {author} {\bibfnamefont {J.~R.}\ \bibnamefont
  {Espinosa}}, \bibinfo {author} {\bibfnamefont {C.}~\bibnamefont {Grojean}},
  \bibinfo {author} {\bibfnamefont {G.}~\bibnamefont {Panico}}, \bibinfo
  {author} {\bibfnamefont {A.}~\bibnamefont {Pomarol}}, \bibinfo {author}
  {\bibfnamefont {O.}~\bibnamefont {Pujol¯Ô}}, \ and\ \bibinfo {author}
  {\bibfnamefont {G.}~\bibnamefont {Servant}},\ }\href@noop {} {\  (\bibinfo
  {year} {2015})},\ \Eprint {http://arxiv.org/abs/1506.09217} {arXiv:1506.09217
  [hep-ph]} \BibitemShut {NoStop}%
%%CITATION = ARXIV:1506.09217;%%
\bibitem [{\citenamefont {Gross}\ \emph {et~al.}(2015)\citenamefont {Gross},
  \citenamefont {Lebedev},\ and\ \citenamefont {Zatta}}]{Gross:2015bea}%
  \BibitemOpen
  \bibfield  {author} {\bibinfo {author} {\bibfnamefont {C.}~\bibnamefont
  {Gross}}, \bibinfo {author} {\bibfnamefont {O.}~\bibnamefont {Lebedev}}, \
  and\ \bibinfo {author} {\bibfnamefont {M.}~\bibnamefont {Zatta}},\
  }\href@noop {} {\  (\bibinfo {year} {2015})},\ \Eprint
  {http://arxiv.org/abs/1506.05106} {arXiv:1506.05106 [hep-ph]} \BibitemShut
  {NoStop}%
%%CITATION = ARXIV:1506.05106;%%
\bibitem [{\citenamefont {Kofman}\ \emph {et~al.}(1997)\citenamefont {Kofman},
  \citenamefont {Linde},\ and\ \citenamefont {Starobinsky}}]{Kofman:1997yn}%
  \BibitemOpen
  \bibfield  {author} {\bibinfo {author} {\bibfnamefont {L.}~\bibnamefont
  {Kofman}}, \bibinfo {author} {\bibfnamefont {A.~D.}\ \bibnamefont {Linde}}, \
  and\ \bibinfo {author} {\bibfnamefont {A.~A.}\ \bibnamefont {Starobinsky}},\
  }\href {\doibase 10.1103/PhysRevD.56.3258} {\bibfield  {journal} {\bibinfo
  {journal} {Phys. Rev.}\ }\textbf {\bibinfo {volume} {D56}},\ \bibinfo {pages}
  {3258} (\bibinfo {year} {1997})},\ \Eprint
  {http://arxiv.org/abs/hep-ph/9704452} {arXiv:hep-ph/9704452 [hep-ph]}
  \BibitemShut {NoStop}%
%%CITATION = HEP-PH/9704452;%%
\bibitem [{\citenamefont {Podolsky}\ \emph {et~al.}(2006)\citenamefont
  {Podolsky}, \citenamefont {Felder}, \citenamefont {Kofman},\ and\
  \citenamefont {Peloso}}]{Podolsky:2005bw}%
  \BibitemOpen
  \bibfield  {author} {\bibinfo {author} {\bibfnamefont {D.~I.}\ \bibnamefont
  {Podolsky}}, \bibinfo {author} {\bibfnamefont {G.~N.}\ \bibnamefont
  {Felder}}, \bibinfo {author} {\bibfnamefont {L.}~\bibnamefont {Kofman}}, \
  and\ \bibinfo {author} {\bibfnamefont {M.}~\bibnamefont {Peloso}},\ }\href
  {\doibase 10.1103/PhysRevD.73.023501} {\bibfield  {journal} {\bibinfo
  {journal} {Phys. Rev.}\ }\textbf {\bibinfo {volume} {D73}},\ \bibinfo {pages}
  {023501} (\bibinfo {year} {2006})},\ \Eprint
  {http://arxiv.org/abs/hep-ph/0507096} {arXiv:hep-ph/0507096 [hep-ph]}
  \BibitemShut {NoStop}%
%%CITATION = HEP-PH/0507096;%%
\bibitem [{\citenamefont {Braden}\ \emph {et~al.}(2010)\citenamefont {Braden},
  \citenamefont {Kofman},\ and\ \citenamefont {Barnaby}}]{Braden:2010wd}%
  \BibitemOpen
  \bibfield  {author} {\bibinfo {author} {\bibfnamefont {J.}~\bibnamefont
  {Braden}}, \bibinfo {author} {\bibfnamefont {L.}~\bibnamefont {Kofman}}, \
  and\ \bibinfo {author} {\bibfnamefont {N.}~\bibnamefont {Barnaby}},\ }\href
  {\doibase 10.1088/1475-7516/2010/07/016} {\bibfield  {journal} {\bibinfo
  {journal} {JCAP}\ }\textbf {\bibinfo {volume} {1007}},\ \bibinfo {pages}
  {016} (\bibinfo {year} {2010})},\ \Eprint {http://arxiv.org/abs/1005.2196}
  {arXiv:1005.2196 [hep-th]} \BibitemShut {NoStop}%
%%CITATION = ARXIV:1005.2196;%%
\bibitem [{\citenamefont {Galloway}\ \emph {et~al.}(2010)\citenamefont
  {Galloway}, \citenamefont {Evans}, \citenamefont {Luty},\ and\ \citenamefont
  {Tacchi}}]{1001.1361}%
  \BibitemOpen
  \bibfield  {author} {\bibinfo {author} {\bibfnamefont {J.}~\bibnamefont
  {Galloway}}, \bibinfo {author} {\bibfnamefont {J.~A.}\ \bibnamefont {Evans}},
  \bibinfo {author} {\bibfnamefont {M.~A.}\ \bibnamefont {Luty}}, \ and\
  \bibinfo {author} {\bibfnamefont {R.~A.}\ \bibnamefont {Tacchi}},\ }\href
  {\doibase 10.1007/JHEP10(2010)086} {\bibfield  {journal} {\bibinfo  {journal}
  {JHEP}\ }\textbf {\bibinfo {volume} {10}},\ \bibinfo {pages} {086} (\bibinfo
  {year} {2010})},\ \Eprint {http://arxiv.org/abs/1001.1361} {arXiv:1001.1361
  [hep-ph]} \BibitemShut {NoStop}%
%%CITATION = ARXIV:1001.1361;%%
\bibitem [{\citenamefont {Gripaios}\ \emph {et~al.}(2009)\citenamefont
  {Gripaios}, \citenamefont {Pomarol}, \citenamefont {Riva},\ and\
  \citenamefont {Serra}}]{0902.1483}%
  \BibitemOpen
  \bibfield  {author} {\bibinfo {author} {\bibfnamefont {B.}~\bibnamefont
  {Gripaios}}, \bibinfo {author} {\bibfnamefont {A.}~\bibnamefont {Pomarol}},
  \bibinfo {author} {\bibfnamefont {F.}~\bibnamefont {Riva}}, \ and\ \bibinfo
  {author} {\bibfnamefont {J.}~\bibnamefont {Serra}},\ }\href {\doibase
  10.1088/1126-6708/2009/04/070} {\bibfield  {journal} {\bibinfo  {journal}
  {JHEP}\ }\textbf {\bibinfo {volume} {04}},\ \bibinfo {pages} {070} (\bibinfo
  {year} {2009})},\ \Eprint {http://arxiv.org/abs/0902.1483} {arXiv:0902.1483
  [hep-ph]} \BibitemShut {NoStop}%
%%CITATION = ARXIV:0902.1483;%%
\bibitem [{\citenamefont {Rubakov}\ and\ \citenamefont
  {Shaposhnikov}(1983)}]{Rubakov:1983bb}%
  \BibitemOpen
  \bibfield  {author} {\bibinfo {author} {\bibfnamefont {V.~A.}\ \bibnamefont
  {Rubakov}}\ and\ \bibinfo {author} {\bibfnamefont {M.~E.}\ \bibnamefont
  {Shaposhnikov}},\ }\href {\doibase 10.1016/0370-2693(83)91253-4} {\bibfield
  {journal} {\bibinfo  {journal} {Phys. Lett.}\ }\textbf {\bibinfo {volume}
  {B125}},\ \bibinfo {pages} {136} (\bibinfo {year} {1983})}\BibitemShut
  {NoStop}%
%%CITATION = PHLTA,B125,136;%%
\bibitem [{\citenamefont {Barbieri}\ and\ \citenamefont
  {Giudice}(1988)}]{Barbieri:1987fn}%
  \BibitemOpen
  \bibfield  {author} {\bibinfo {author} {\bibfnamefont {R.}~\bibnamefont
  {Barbieri}}\ and\ \bibinfo {author} {\bibfnamefont {G.~F.}\ \bibnamefont
  {Giudice}},\ }\href {\doibase 10.1016/0550-3213(88)90171-X} {\bibfield
  {journal} {\bibinfo  {journal} {Nucl. Phys.}\ }\textbf {\bibinfo {volume}
  {B306}},\ \bibinfo {pages} {63} (\bibinfo {year} {1988})}\BibitemShut
  {NoStop}%
%%CITATION = NUPHA,B306,63;%%
\bibitem [{\citenamefont {Allahverdi}\ \emph {et~al.}(2010)\citenamefont
  {Allahverdi}, \citenamefont {Brandenberger}, \citenamefont {Cyr-Racine},\
  and\ \citenamefont {Mazumdar}}]{Allahverdi:2010xz}%
  \BibitemOpen
  \bibfield  {author} {\bibinfo {author} {\bibfnamefont {R.}~\bibnamefont
  {Allahverdi}}, \bibinfo {author} {\bibfnamefont {R.}~\bibnamefont
  {Brandenberger}}, \bibinfo {author} {\bibfnamefont {F.-Y.}\ \bibnamefont
  {Cyr-Racine}}, \ and\ \bibinfo {author} {\bibfnamefont {A.}~\bibnamefont
  {Mazumdar}},\ }\href {\doibase 10.1146/annurev.nucl.012809.104511} {\bibfield
   {journal} {\bibinfo  {journal} {Ann.Rev.Nucl.Part.Sci.}\ }\textbf {\bibinfo
  {volume} {60}},\ \bibinfo {pages} {27} (\bibinfo {year} {2010})},\ \Eprint
  {http://arxiv.org/abs/1001.2600} {arXiv:1001.2600 [hep-th]} \BibitemShut
  {NoStop}%
%%CITATION = ARXIV:1001.2600;%%
\bibitem [{\citenamefont {Traschen}\ and\ \citenamefont
  {Brandenberger}(1990)}]{Traschen:1990sw}%
  \BibitemOpen
  \bibfield  {author} {\bibinfo {author} {\bibfnamefont {J.~H.}\ \bibnamefont
  {Traschen}}\ and\ \bibinfo {author} {\bibfnamefont {R.~H.}\ \bibnamefont
  {Brandenberger}},\ }\href {\doibase 10.1103/PhysRevD.42.2491} {\bibfield
  {journal} {\bibinfo  {journal} {Phys.Rev.}\ }\textbf {\bibinfo {volume}
  {D42}},\ \bibinfo {pages} {2491} (\bibinfo {year} {1990})}\BibitemShut
  {NoStop}%
%%CITATION = PHRVA,D42,2491;%%
\bibitem [{\citenamefont {Shtanov}\ \emph {et~al.}(1995)\citenamefont
  {Shtanov}, \citenamefont {Traschen},\ and\ \citenamefont
  {Brandenberger}}]{Shtanov:1994ce}%
  \BibitemOpen
  \bibfield  {author} {\bibinfo {author} {\bibfnamefont {Y.}~\bibnamefont
  {Shtanov}}, \bibinfo {author} {\bibfnamefont {J.~H.}\ \bibnamefont
  {Traschen}}, \ and\ \bibinfo {author} {\bibfnamefont {R.~H.}\ \bibnamefont
  {Brandenberger}},\ }\href {\doibase 10.1103/PhysRevD.51.5438} {\bibfield
  {journal} {\bibinfo  {journal} {Phys. Rev.}\ }\textbf {\bibinfo {volume}
  {D51}},\ \bibinfo {pages} {5438} (\bibinfo {year} {1995})},\ \Eprint
  {http://arxiv.org/abs/hep-ph/9407247} {arXiv:hep-ph/9407247 [hep-ph]}
  \BibitemShut {NoStop}%
%%CITATION = HEP-PH/9407247;%%
\bibitem [{\citenamefont {Linde}\ \emph {et~al.}(2013)\citenamefont {Linde},
  \citenamefont {Mooij},\ and\ \citenamefont {Pajer}}]{Linde:2012bt}%
  \BibitemOpen
  \bibfield  {author} {\bibinfo {author} {\bibfnamefont {A.}~\bibnamefont
  {Linde}}, \bibinfo {author} {\bibfnamefont {S.}~\bibnamefont {Mooij}}, \ and\
  \bibinfo {author} {\bibfnamefont {E.}~\bibnamefont {Pajer}},\ }\href
  {\doibase 10.1103/PhysRevD.87.103506} {\bibfield  {journal} {\bibinfo
  {journal} {Phys. Rev.}\ }\textbf {\bibinfo {volume} {D87}},\ \bibinfo {pages}
  {103506} (\bibinfo {year} {2013})},\ \Eprint {http://arxiv.org/abs/1212.1693}
  {arXiv:1212.1693 [hep-th]} \BibitemShut {NoStop}%
%%CITATION = ARXIV:1212.1693;%%
\bibitem [{\citenamefont {Adshead}\ \emph {et~al.}(2015)\citenamefont
  {Adshead}, \citenamefont {Giblin}, \citenamefont {Scully},\ and\
  \citenamefont {Sfakianakis}}]{Adshead:2015pva}%
  \BibitemOpen
  \bibfield  {author} {\bibinfo {author} {\bibfnamefont {P.}~\bibnamefont
  {Adshead}}, \bibinfo {author} {\bibfnamefont {J.~T.}\ \bibnamefont {Giblin}},
  \bibinfo {author} {\bibfnamefont {T.~R.}\ \bibnamefont {Scully}}, \ and\
  \bibinfo {author} {\bibfnamefont {E.~I.}\ \bibnamefont {Sfakianakis}},\
  }\href@noop {} {\  (\bibinfo {year} {2015})},\ \Eprint
  {http://arxiv.org/abs/1502.06506} {arXiv:1502.06506 [astro-ph.CO]}
  \BibitemShut {NoStop}%
%%CITATION = ARXIV:1502.06506;%%
\bibitem [{\citenamefont {Kolb}\ \emph {et~al.}(2003)\citenamefont {Kolb},
  \citenamefont {Notari},\ and\ \citenamefont {Riotto}}]{Kolb:2003ke}%
  \BibitemOpen
  \bibfield  {author} {\bibinfo {author} {\bibfnamefont {E.~W.}\ \bibnamefont
  {Kolb}}, \bibinfo {author} {\bibfnamefont {A.}~\bibnamefont {Notari}}, \ and\
  \bibinfo {author} {\bibfnamefont {A.}~\bibnamefont {Riotto}},\ }\href
  {\doibase 10.1103/PhysRevD.68.123505} {\bibfield  {journal} {\bibinfo
  {journal} {Phys.Rev.}\ }\textbf {\bibinfo {volume} {D68}},\ \bibinfo {pages}
  {123505} (\bibinfo {year} {2003})},\ \Eprint
  {http://arxiv.org/abs/hep-ph/0307241} {arXiv:hep-ph/0307241 [hep-ph]}
  \BibitemShut {NoStop}%
%%CITATION = HEP-PH/0307241;%%
\bibitem [{\citenamefont {Drewes}(2014)}]{Drewes:2014pfa}%
  \BibitemOpen
  \bibfield  {author} {\bibinfo {author} {\bibfnamefont {M.}~\bibnamefont
  {Drewes}},\ }\href {\doibase 10.1088/1475-7516/2014/11/020} {\bibfield
  {journal} {\bibinfo  {journal} {JCAP}\ }\textbf {\bibinfo {volume} {1411}},\
  \bibinfo {pages} {020} (\bibinfo {year} {2014})},\ \Eprint
  {http://arxiv.org/abs/1406.6243} {arXiv:1406.6243 [hep-ph]} \BibitemShut
  {NoStop}%
%%CITATION = ARXIV:1406.6243;%%
\bibitem [{\citenamefont {Drewes}\ and\ \citenamefont
  {Kang}(2013)}]{Drewes:2013iaa}%
  \BibitemOpen
  \bibfield  {author} {\bibinfo {author} {\bibfnamefont {M.}~\bibnamefont
  {Drewes}}\ and\ \bibinfo {author} {\bibfnamefont {J.~U.}\ \bibnamefont
  {Kang}},\ }\href {\doibase 10.1016/j.nuclphysb.2013.07.009,
  10.1016/j.nuclphysb.2014.09.008} {\bibfield  {journal} {\bibinfo  {journal}
  {Nucl.Phys.}\ }\textbf {\bibinfo {volume} {B875}},\ \bibinfo {pages} {315}
  (\bibinfo {year} {2013})},\ \Eprint {http://arxiv.org/abs/1305.0267}
  {arXiv:1305.0267 [hep-ph]} \BibitemShut {NoStop}%
%%CITATION = ARXIV:1305.0267;%%
\bibitem [{\citenamefont {Kawasaki}\ \emph {et~al.}(1999)\citenamefont
  {Kawasaki}, \citenamefont {Kohri},\ and\ \citenamefont
  {Sugiyama}}]{Kawasaki:1999na}%
  \BibitemOpen
  \bibfield  {author} {\bibinfo {author} {\bibfnamefont {M.}~\bibnamefont
  {Kawasaki}}, \bibinfo {author} {\bibfnamefont {K.}~\bibnamefont {Kohri}}, \
  and\ \bibinfo {author} {\bibfnamefont {N.}~\bibnamefont {Sugiyama}},\ }\href
  {\doibase 10.1103/PhysRevLett.82.4168} {\bibfield  {journal} {\bibinfo
  {journal} {Phys. Rev. Lett.}\ }\textbf {\bibinfo {volume} {82}},\ \bibinfo
  {pages} {4168} (\bibinfo {year} {1999})},\ \Eprint
  {http://arxiv.org/abs/astro-ph/9811437} {arXiv:astro-ph/9811437 [astro-ph]}
  \BibitemShut {NoStop}%
%%CITATION = ASTRO-PH/9811437;%%
\bibitem [{\citenamefont {Ichikawa}\ \emph {et~al.}(2005)\citenamefont
  {Ichikawa}, \citenamefont {Kawasaki},\ and\ \citenamefont
  {Takahashi}}]{Ichikawa:2005vw}%
  \BibitemOpen
  \bibfield  {author} {\bibinfo {author} {\bibfnamefont {K.}~\bibnamefont
  {Ichikawa}}, \bibinfo {author} {\bibfnamefont {M.}~\bibnamefont {Kawasaki}},
  \ and\ \bibinfo {author} {\bibfnamefont {F.}~\bibnamefont {Takahashi}},\
  }\href {\doibase 10.1103/PhysRevD.72.043522} {\bibfield  {journal} {\bibinfo
  {journal} {Phys. Rev.}\ }\textbf {\bibinfo {volume} {D72}},\ \bibinfo {pages}
  {043522} (\bibinfo {year} {2005})},\ \Eprint
  {http://arxiv.org/abs/astro-ph/0505395} {arXiv:astro-ph/0505395 [astro-ph]}
  \BibitemShut {NoStop}%
%%CITATION = ASTRO-PH/0505395;%%
\bibitem [{\citenamefont {Martin}\ \emph {et~al.}(2015)\citenamefont {Martin},
  \citenamefont {Ringeval},\ and\ \citenamefont {Vennin}}]{Martin:2014nya}%
  \BibitemOpen
  \bibfield  {author} {\bibinfo {author} {\bibfnamefont {J.}~\bibnamefont
  {Martin}}, \bibinfo {author} {\bibfnamefont {C.}~\bibnamefont {Ringeval}}, \
  and\ \bibinfo {author} {\bibfnamefont {V.}~\bibnamefont {Vennin}},\ }\href
  {\doibase 10.1103/PhysRevLett.114.081303} {\bibfield  {journal} {\bibinfo
  {journal} {Phys. Rev. Lett.}\ }\textbf {\bibinfo {volume} {114}},\ \bibinfo
  {pages} {081303} (\bibinfo {year} {2015})},\ \Eprint
  {http://arxiv.org/abs/1410.7958} {arXiv:1410.7958 [astro-ph.CO]} \BibitemShut
  {NoStop}%
%%CITATION = ARXIV:1410.7958;%%
\bibitem [{\citenamefont {Hannestad}(2004)}]{Hannestad:2004px}%
  \BibitemOpen
  \bibfield  {author} {\bibinfo {author} {\bibfnamefont {S.}~\bibnamefont
  {Hannestad}},\ }\href {\doibase 10.1103/PhysRevD.70.043506} {\bibfield
  {journal} {\bibinfo  {journal} {Phys. Rev.}\ }\textbf {\bibinfo {volume}
  {D70}},\ \bibinfo {pages} {043506} (\bibinfo {year} {2004})},\ \Eprint
  {http://arxiv.org/abs/astro-ph/0403291} {arXiv:astro-ph/0403291 [astro-ph]}
  \BibitemShut {NoStop}%
%%CITATION = ASTRO-PH/0403291;%%
\bibitem [{\citenamefont {Cohen}\ \emph {et~al.}(1993)\citenamefont {Cohen},
  \citenamefont {Kaplan},\ and\ \citenamefont {Nelson}}]{Cohen:1993nk}%
  \BibitemOpen
  \bibfield  {author} {\bibinfo {author} {\bibfnamefont {A.~G.}\ \bibnamefont
  {Cohen}}, \bibinfo {author} {\bibfnamefont {D.~B.}\ \bibnamefont {Kaplan}}, \
  and\ \bibinfo {author} {\bibfnamefont {A.~E.}\ \bibnamefont {Nelson}},\
  }\href {\doibase 10.1146/annurev.ns.43.120193.000331} {\bibfield  {journal}
  {\bibinfo  {journal} {Ann. Rev. Nucl. Part. Sci.}\ }\textbf {\bibinfo
  {volume} {43}},\ \bibinfo {pages} {27} (\bibinfo {year} {1993})},\ \Eprint
  {http://arxiv.org/abs/hep-ph/9302210} {arXiv:hep-ph/9302210 [hep-ph]}
  \BibitemShut {NoStop}%
%%CITATION = HEP-PH/9302210;%%
\bibitem [{\citenamefont {Espinosa}\ \emph {et~al.}(2012)\citenamefont
  {Espinosa}, \citenamefont {Gripaios}, \citenamefont {Konstandin},\ and\
  \citenamefont {Riva}}]{Espinosa:2011eu}%
  \BibitemOpen
  \bibfield  {author} {\bibinfo {author} {\bibfnamefont {J.~R.}\ \bibnamefont
  {Espinosa}}, \bibinfo {author} {\bibfnamefont {B.}~\bibnamefont {Gripaios}},
  \bibinfo {author} {\bibfnamefont {T.}~\bibnamefont {Konstandin}}, \ and\
  \bibinfo {author} {\bibfnamefont {F.}~\bibnamefont {Riva}},\ }\href {\doibase
  10.1088/1475-7516/2012/01/012} {\bibfield  {journal} {\bibinfo  {journal}
  {JCAP}\ }\textbf {\bibinfo {volume} {1201}},\ \bibinfo {pages} {012}
  (\bibinfo {year} {2012})},\ \Eprint {http://arxiv.org/abs/1110.2876}
  {arXiv:1110.2876 [hep-ph]} \BibitemShut {NoStop}%
%%CITATION = ARXIV:1110.2876;%%
\bibitem [{\citenamefont {Gorbahn}\ \emph {et~al.}(2015)\citenamefont
  {Gorbahn}, \citenamefont {No},\ and\ \citenamefont {Sanz}}]{Gorbahn:2015gxa}%
  \BibitemOpen
  \bibfield  {author} {\bibinfo {author} {\bibfnamefont {M.}~\bibnamefont
  {Gorbahn}}, \bibinfo {author} {\bibfnamefont {J.~M.}\ \bibnamefont {No}}, \
  and\ \bibinfo {author} {\bibfnamefont {V.}~\bibnamefont {Sanz}},\ }\href@noop
  {} {\  (\bibinfo {year} {2015})},\ \Eprint {http://arxiv.org/abs/1502.07352}
  {arXiv:1502.07352 [hep-ph]} \BibitemShut {NoStop}%
%%CITATION = ARXIV:1502.07352;%%
\bibitem [{\citenamefont {Aad}\ \emph {et~al.}(2014)\citenamefont {Aad} \emph
  {et~al.}}]{Aad:2014eha}%
  \BibitemOpen
  \bibfield  {author} {\bibinfo {author} {\bibfnamefont {G.}~\bibnamefont
  {Aad}} \emph {et~al.} (\bibinfo {collaboration} {ATLAS}),\ }\href {\doibase
  10.1103/PhysRevD.90.112015} {\bibfield  {journal} {\bibinfo  {journal} {Phys.
  Rev.}\ }\textbf {\bibinfo {volume} {D90}},\ \bibinfo {pages} {112015}
  (\bibinfo {year} {2014})},\ \Eprint {http://arxiv.org/abs/1408.7084}
  {arXiv:1408.7084 [hep-ex]} \BibitemShut {NoStop}%
%%CITATION = ARXIV:1408.7084;%%
\bibitem [{\citenamefont {Khachatryan}\ \emph {et~al.}(2014)\citenamefont
  {Khachatryan} \emph {et~al.}}]{Khachatryan:2014ira}%
  \BibitemOpen
  \bibfield  {author} {\bibinfo {author} {\bibfnamefont {V.}~\bibnamefont
  {Khachatryan}} \emph {et~al.} (\bibinfo {collaboration} {CMS}),\ }\href
  {\doibase 10.1140/epjc/s10052-014-3076-z} {\bibfield  {journal} {\bibinfo
  {journal} {Eur. Phys. J.}\ }\textbf {\bibinfo {volume} {C74}},\ \bibinfo
  {pages} {3076} (\bibinfo {year} {2014})},\ \Eprint
  {http://arxiv.org/abs/1407.0558} {arXiv:1407.0558 [hep-ex]} \BibitemShut
  {NoStop}%
%%CITATION = ARXIV:1407.0558;%%
\bibitem [{\citenamefont {Aad}\ \emph {et~al.}(2015)\citenamefont {Aad} \emph
  {et~al.}}]{Aad:2014eva}%
  \BibitemOpen
  \bibfield  {author} {\bibinfo {author} {\bibfnamefont {G.}~\bibnamefont
  {Aad}} \emph {et~al.} (\bibinfo {collaboration} {ATLAS}),\ }\href {\doibase
  10.1103/PhysRevD.91.012006} {\bibfield  {journal} {\bibinfo  {journal} {Phys.
  Rev.}\ }\textbf {\bibinfo {volume} {D91}},\ \bibinfo {pages} {012006}
  (\bibinfo {year} {2015})},\ \Eprint {http://arxiv.org/abs/1408.5191}
  {arXiv:1408.5191 [hep-ex]} \BibitemShut {NoStop}%
%%CITATION = ARXIV:1408.5191;%%
\bibitem [{\citenamefont {Chatrchyan}\ \emph
  {et~al.}(2014{\natexlab{a}})\citenamefont {Chatrchyan} \emph
  {et~al.}}]{Chatrchyan:2013mxa}%
  \BibitemOpen
  \bibfield  {author} {\bibinfo {author} {\bibfnamefont {S.}~\bibnamefont
  {Chatrchyan}} \emph {et~al.} (\bibinfo {collaboration} {CMS}),\ }\href
  {\doibase 10.1103/PhysRevD.89.092007} {\bibfield  {journal} {\bibinfo
  {journal} {Phys. Rev.}\ }\textbf {\bibinfo {volume} {D89}},\ \bibinfo {pages}
  {092007} (\bibinfo {year} {2014}{\natexlab{a}})},\ \Eprint
  {http://arxiv.org/abs/1312.5353} {arXiv:1312.5353 [hep-ex]} \BibitemShut
  {NoStop}%
%%CITATION = ARXIV:1312.5353;%%
\bibitem [{1229952()}]{ATLAS:2013wla}%
  \BibitemOpen
  \bibfield  {author} {1229952,\ }\href@noop {} {\  (\bibinfo {year}
  {2013})}\BibitemShut {NoStop}%
%%CITATION = ATLAS-CONF-2013-030 ETC.;%%
\bibitem [{\citenamefont {Chatrchyan}\ \emph
  {et~al.}(2014{\natexlab{b}})\citenamefont {Chatrchyan} \emph
  {et~al.}}]{Chatrchyan:2013iaa}%
  \BibitemOpen
  \bibfield  {author} {\bibinfo {author} {\bibfnamefont {S.}~\bibnamefont
  {Chatrchyan}} \emph {et~al.} (\bibinfo {collaboration} {CMS}),\ }\href
  {\doibase 10.1007/JHEP01(2014)096} {\bibfield  {journal} {\bibinfo  {journal}
  {JHEP}\ }\textbf {\bibinfo {volume} {01}},\ \bibinfo {pages} {096} (\bibinfo
  {year} {2014}{\natexlab{b}})},\ \Eprint {http://arxiv.org/abs/1312.1129}
  {arXiv:1312.1129 [hep-ex]} \BibitemShut {NoStop}%
%%CITATION = ARXIV:1312.1129;%%
\bibitem [{\citenamefont
  {collaboration}(2013{\natexlab{a}})}]{TheATLAScollaboration:2013lia}%
  \BibitemOpen
  \bibfield  {author} {\bibinfo {author} {\bibfnamefont {T.~A.}\ \bibnamefont
  {collaboration}} (\bibinfo {collaboration} {ATLAS}),\ }\href@noop {} {\
  (\bibinfo {year} {2013}{\natexlab{a}})}\BibitemShut {NoStop}%
%%CITATION = ATLAS-CONF-2013-079;%%
\bibitem [{\citenamefont {Chatrchyan}\ \emph
  {et~al.}(2014{\natexlab{c}})\citenamefont {Chatrchyan} \emph
  {et~al.}}]{Chatrchyan:2013zna}%
  \BibitemOpen
  \bibfield  {author} {\bibinfo {author} {\bibfnamefont {S.}~\bibnamefont
  {Chatrchyan}} \emph {et~al.} (\bibinfo {collaboration} {CMS}),\ }\href
  {\doibase 10.1103/PhysRevD.89.012003} {\bibfield  {journal} {\bibinfo
  {journal} {Phys. Rev.}\ }\textbf {\bibinfo {volume} {D89}},\ \bibinfo {pages}
  {012003} (\bibinfo {year} {2014}{\natexlab{c}})},\ \Eprint
  {http://arxiv.org/abs/1310.3687} {arXiv:1310.3687 [hep-ex]} \BibitemShut
  {NoStop}%
%%CITATION = ARXIV:1310.3687;%%
\bibitem [{\citenamefont {collaboration}(2013{\natexlab{b}})}]{ATLAS_tau}%
  \BibitemOpen
  \bibfield  {author} {\bibinfo {author} {\bibfnamefont {T.~A.}\ \bibnamefont
  {collaboration}} (\bibinfo {collaboration} {ATLAS}),\ }\href@noop {} {\
  (\bibinfo {year} {2013}{\natexlab{b}})}\BibitemShut {NoStop}%
%%CITATION = ATLAS-CONF-2013-108;%%
\bibitem [{\citenamefont {Chatrchyan}\ \emph
  {et~al.}(2014{\natexlab{d}})\citenamefont {Chatrchyan} \emph
  {et~al.}}]{Chatrchyan:2014nva}%
  \BibitemOpen
  \bibfield  {author} {\bibinfo {author} {\bibfnamefont {S.}~\bibnamefont
  {Chatrchyan}} \emph {et~al.} (\bibinfo {collaboration} {CMS}),\ }\href
  {\doibase 10.1007/JHEP05(2014)104} {\bibfield  {journal} {\bibinfo  {journal}
  {JHEP}\ }\textbf {\bibinfo {volume} {05}},\ \bibinfo {pages} {104} (\bibinfo
  {year} {2014}{\natexlab{d}})},\ \Eprint {http://arxiv.org/abs/1401.5041}
  {arXiv:1401.5041 [hep-ex]} \BibitemShut {NoStop}%
%%CITATION = ARXIV:1401.5041;%%
\bibitem [{\citenamefont {Baak}\ \emph {et~al.}(2014)\citenamefont {Baak},
  \citenamefont {Cth}, \citenamefont {Haller}, \citenamefont {Hoecker},
  \citenamefont {Kogler}, \citenamefont {Mnig}, \citenamefont {Schott},\ and\
  \citenamefont {Stelzer}}]{Baak:2014ora}%
  \BibitemOpen
  \bibfield  {author} {\bibinfo {author} {\bibfnamefont {M.}~\bibnamefont
  {Baak}}, \bibinfo {author} {\bibfnamefont {J.}~\bibnamefont {Cth}}, \bibinfo
  {author} {\bibfnamefont {J.}~\bibnamefont {Haller}}, \bibinfo {author}
  {\bibfnamefont {A.}~\bibnamefont {Hoecker}}, \bibinfo {author} {\bibfnamefont
  {R.}~\bibnamefont {Kogler}}, \bibinfo {author} {\bibfnamefont
  {K.}~\bibnamefont {Mnig}}, \bibinfo {author} {\bibfnamefont {M.}~\bibnamefont
  {Schott}}, \ and\ \bibinfo {author} {\bibfnamefont {J.}~\bibnamefont
  {Stelzer}} (\bibinfo {collaboration} {Gfitter Group}),\ }\href {\doibase
  10.1140/epjc/s10052-014-3046-5} {\bibfield  {journal} {\bibinfo  {journal}
  {Eur. Phys. J.}\ }\textbf {\bibinfo {volume} {C74}},\ \bibinfo {pages} {3046}
  (\bibinfo {year} {2014})},\ \Eprint {http://arxiv.org/abs/1407.3792}
  {arXiv:1407.3792 [hep-ph]} \BibitemShut {NoStop}%
%%CITATION = ARXIV:1407.3792;%%
\bibitem [{\citenamefont {Fan}\ \emph {et~al.}(2014)\citenamefont {Fan},
  \citenamefont {Reece},\ and\ \citenamefont {Wang}}]{Fan:2014vta}%
  \BibitemOpen
  \bibfield  {author} {\bibinfo {author} {\bibfnamefont {J.}~\bibnamefont
  {Fan}}, \bibinfo {author} {\bibfnamefont {M.}~\bibnamefont {Reece}}, \ and\
  \bibinfo {author} {\bibfnamefont {L.-T.}\ \bibnamefont {Wang}},\ }\href@noop
  {} {\  (\bibinfo {year} {2014})},\ \Eprint {http://arxiv.org/abs/1411.1054}
  {arXiv:1411.1054 [hep-ph]} \BibitemShut {NoStop}%
%%CITATION = ARXIV:1411.1054;%%
\bibitem [{\citenamefont {Hagiwara}\ \emph {et~al.}(1994)\citenamefont
  {Hagiwara}, \citenamefont {Matsumoto}, \citenamefont {Haidt},\ and\
  \citenamefont {Kim}}]{Hagiwara:1994pw}%
  \BibitemOpen
  \bibfield  {author} {\bibinfo {author} {\bibfnamefont {K.}~\bibnamefont
  {Hagiwara}}, \bibinfo {author} {\bibfnamefont {S.}~\bibnamefont {Matsumoto}},
  \bibinfo {author} {\bibfnamefont {D.}~\bibnamefont {Haidt}}, \ and\ \bibinfo
  {author} {\bibfnamefont {C.~S.}\ \bibnamefont {Kim}},\ }\href {\doibase
  10.1007/BF01957770} {\bibfield  {journal} {\bibinfo  {journal} {Z. Phys.}\
  }\textbf {\bibinfo {volume} {C64}},\ \bibinfo {pages} {559} (\bibinfo {year}
  {1994})},\ \bibinfo {note} {[Erratum: Z. Phys.C68,352(1995)]},\ \Eprint
  {http://arxiv.org/abs/hep-ph/9409380} {arXiv:hep-ph/9409380 [hep-ph]}
  \BibitemShut {NoStop}%
%%CITATION = HEP-PH/9409380;%%
\bibitem [{\citenamefont {Asner}\ \emph {et~al.}(2013)\citenamefont {Asner}
  \emph {et~al.}}]{Asner:2013psa}%
  \BibitemOpen
  \bibfield  {author} {\bibinfo {author} {\bibfnamefont {D.~M.}\ \bibnamefont
  {Asner}} \emph {et~al.},\ }in\ \href
  {https://inspirehep.net/record/1256491/files/arXiv:1310.0763.pdf} {\emph
  {\bibinfo {booktitle} {{Community Summer Study 2013: Snowmass on the
  Mississippi (CSS2013) Minneapolis, MN, USA, July 29-August 6, 2013}}}}\
  (\bibinfo {year} {2013})\ \Eprint {http://arxiv.org/abs/1310.0763}
  {arXiv:1310.0763 [hep-ph]} \BibitemShut {NoStop}%
%%CITATION = ARXIV:1310.0763;%%
\bibitem [{\citenamefont {Bicer}\ \emph {et~al.}(2014)\citenamefont {Bicer}
  \emph {et~al.}}]{Gomez-Ceballos:2013zzn}%
  \BibitemOpen
  \bibfield  {author} {\bibinfo {author} {\bibfnamefont {M.}~\bibnamefont
  {Bicer}} \emph {et~al.} (\bibinfo {collaboration} {TLEP Design Study Working
  Group}),\ }\bibfield  {booktitle} {\emph {\bibinfo {booktitle} {{Community
  Summer Study 2013: Snowmass on the Mississippi (CSS2013) Minneapolis, MN,
  USA, July 29-August 6, 2013}}},\ }\href {\doibase 10.1007/JHEP01(2014)164}
  {\bibfield  {journal} {\bibinfo  {journal} {JHEP}\ }\textbf {\bibinfo
  {volume} {01}},\ \bibinfo {pages} {164} (\bibinfo {year} {2014})},\ \Eprint
  {http://arxiv.org/abs/1308.6176} {arXiv:1308.6176 [hep-ex]} \BibitemShut
  {NoStop}%
%%CITATION = ARXIV:1308.6176;%%
\bibitem [{\citenamefont {Redi}\ and\ \citenamefont
  {Tesi}(2012)}]{Redi:2012ha}%
  \BibitemOpen
  \bibfield  {author} {\bibinfo {author} {\bibfnamefont {M.}~\bibnamefont
  {Redi}}\ and\ \bibinfo {author} {\bibfnamefont {A.}~\bibnamefont {Tesi}},\
  }\href {\doibase 10.1007/JHEP10(2012)166} {\bibfield  {journal} {\bibinfo
  {journal} {JHEP}\ }\textbf {\bibinfo {volume} {10}},\ \bibinfo {pages} {166}
  (\bibinfo {year} {2012})},\ \Eprint {http://arxiv.org/abs/1205.0232}
  {arXiv:1205.0232 [hep-ph]} \BibitemShut {NoStop}%
%%CITATION = ARXIV:1205.0232;%%
\end{thebibliography}%

 \end{document}